\documentclass[aps,prb,twocolumn,groupedaddress]{revtex4-1}
\usepackage[dvips]{graphicx}
\usepackage{braket}
\begin{document}

\renewcommand{\vec}[1]{\mathbf{#1}}




\title{\textit{Ab initio} computational study on the lattice thermal conductivity of Zintl clathrates [Si$_{19}$P$_{4}$]Cl$_{4}$ and Na$_{4}$[Al$_{4}$Si$_{19}$]}


\author{Ville J. H\"{a}rk\"{o}nen}
\email[]{ville.j.harkonen@jyu.fi}
\affiliation{Department of Chemistry, University of Jyv\"{a}skyl\"{a}, PO Box 35, FI-40014, Finland}
\author{Antti J. Karttunen}
\email[]{antti.j.karttunen@iki.fi}
\affiliation{Department of Chemistry, Aalto University, FI-00076 Espoo, Finland.}


\date{\today}

\begin{abstract}
The lattice thermal conductivity of silicon clathrate framework Si$_{23}$ and two Zintl clathrates, [Si$_{19}$P$_{4}$]Cl$_{4}$ and Na$_{4}$[Al$_{4}$Si$_{19}$], is investigated by using an iterative solution of the linearized Boltzmann transport equation (BTE) in conjuntion with \textit{ab initio} lattice dynamical techniques. At 300 K, the lattice thermal conductivities for Si$_{23}$, [Si$_{19}$P$_{4}$]Cl$_{4}$, and Na$_{4}$[Al$_{4}$Si$_{19}$] were found to be 43 W/(m K), 25 W/(m K), and 2 W/(m K), respectively. In the case of Na$_{4}$[Al$_{4}$Si$_{19}$], the order-of-magnitude reduction in the lattice thermal conductivity was found to be mostly due to relaxation times and group velocities differing from Si$_{23}$ and [Si$_{19}$P$_{4}$]Cl$_{4}$. The difference in the relaxation times and group velocities arises primarily due to the phonon spectrum at low frequencies, resulting eventually from the differences in the second-order interatomic force constants (IFCs). The obtained third-order IFCs were rather similar for all materials considered here.
\end{abstract}


\maketitle

\section{Introduction}
\label{cha:Introduction}
The minimization of the lattice thermal conductivity is usually a desired feature when higher thermoelectric efficiency is pursued \cite{Ioffe-SemicondThermoelements-1957,Slack-crtThermoelectrics-1995}. There are several different classes of materials that appear to be promising for the thermoelectric applications, one example of such materials being the Zintl clathrates, also known as semiconducting clathrates \cite{Kasper-24121965.science.clath.1965,Nolas-Ge-clath-thermoelect-1998,Kovnir-0036-021X-73-9-R06.clath.review-2004,Karttunen-Structuralprinc-2010,Takabatake-RevModPhys.86.669-2014,Norouzzadeh-GiantPowerFactorClatVIII-2014}. The lattice thermal conductivity of the Zintl clathrates has been studied rather intensively in the past two decades \cite{Cohn-GlasslikeHeatCond-PhysRevLett.82.779-1999,Tse-TcondClath-PhysRevLett.85.114-2000,Dong-LatCondGeClath-PhysRevLett.86.2361-2001,Nolas-ThermalCondSi136-2003,Bentien-TcondClath-PhysRevB.69.045107-2004,Avila-GlassLikeTCond-PhysRevB.74.125109-2006,Suekuni-CageSize-PhysRevB.75.195210-2007,Takasu-TypeIclathTCond-PhysRevLett.100.165503-2008,Christensen-AvoidedCrossing-2008,Avila-ba8ga16sn30TCond-2008,Christensen-ClathrateGuest-2009,Christensen.et.al.-2010-B916400F,Candolfi-TransportClathrates-2011,Euchner-RatlingClath-PhysRevB.86.224303-2012,Fulmer-TcondClathrate-2013,He-NanostructuredClath-2014,Pailhes-Localization-PhysRevLett.113.025506-2014,Castillo-Clathrate-2015,Tadano-ImpactOfRattlers-PhysRevLett.114.095501-2015,Kishimoto-ThermoelectricClathII-2015,Norouzzadeh-CTEpaper-2016,Madsen-LatCondQHA-2016} and several mechanisms have been proposed to explain the reduced lattice thermal conductivity values in these materials. In particular, lattice thermal conductivity values as low as $\sim 1$ W/m K at 150 K have been obtained experimentally for some silicon clathrates \cite{Takasu-TypeIclathTCond-PhysRevLett.100.165503-2008,Takabatake-RevModPhys.86.669-2014} (single crystal samples). Recent experimental and computational studies have given rise to different points of view for the reasons behind the reduction of the lattice thermal conductivity in various clathrates \cite{Pailhes-Localization-PhysRevLett.113.025506-2014,Tadano-ImpactOfRattlers-PhysRevLett.114.095501-2015}. In Ref. \cite{Pailhes-Localization-PhysRevLett.113.025506-2014} it was concluded that the reduction of the lattice thermal conductivity of the Ba$_{8}$Si$_{46}$ clathrate is mostly due to the harmonic phonon spectrum, while in Ref. \cite{Tadano-ImpactOfRattlers-PhysRevLett.114.095501-2015} for the clathrate Ba$_{8}$Ga$_{16}$Ge$_{30}$ the reduction was suggested to arise mainly from rather short relaxation times (RTs). This work brings further perspectives on these issues by using computational techniques applied on two different types of Zintl clathrates.

In this work, we study the lattice thermal conductivity of the silicon clathrate framework Si$_{23}$ (sometimes denoted as VIII or Si$_{46}$-VIII) and two hypothetical Zintl clathrate structures [Si$_{19}$P$_{4}$]Cl$_{4}$ and Na$_{4}$[Al$_{4}$Si$_{19}$], obtained by adding guest atoms and framework heteroatoms in the Si$_{23}$ structure. All the considered structures possess the same space group symmetry, which facilitates the comparative analysis of these materials since some quantities for the materials are identical. The study is carried out by using the Boltzmann transport equation (BTE) approach implemented in the open source program package ShengBTE. The harmonic phonon eigenvalues and eigenvectors used in the lattice thermal conductivity calculations are obtained by using density functional perturbation theory as implemented in the Quantum Espresso (QE) program package. In particular, the third-order interatomic force constants (IFCs) and quantities within the harmonic approximation are analyzed in order to understand their role in the reduction of lattice thermal conductivity. The results indicate that in the studied structures, the increased anharmonicity can be mostly explained by harmonic quantities and rather low lattice thermal conductivity values do not necessarily indicate exceptionally strong third-order IFCs.

\section{Theory, computational methods and studied structures}
\label{cha:TheoryComputationalMethodsAndStudiedStructures}

\subsection{Lattice dynamics}
\label{LatticeDynamics}
The theory of lattice dynamics discussed in this section has been considered, for instance, in Refs. \cite{Born-Huang-DynamicalTheory-1954,Maradudin-harm-appr-1971,Maradudin-DynamicalPropertiesOfSolids-1974}. The notation used here is the same as in Ref. \cite{Harkonen-Tcond-II-VIII-PhysRevB.93.024307-2016}. In the present approach, one assumes that the lattice Hamiltonian is of the form 
\begin{equation} 
\hat{H} = \hat{H}_{0} + \hat{H}_{A},
\label{eq:LatticeDynamicsEq_1}
\end{equation}
where the harmonic Hamiltonian operator $\hat{H}_{0}$ may be written as
\begin{equation} 
\hat{H}_{0}=\sum_{\lambda}\hbar \omega_{\lambda} \left(\frac{1}{2}+ \hat{a}^{\dagger}_{\lambda} \hat{a}_{\lambda} \right), \quad \lambda \equiv \vec{q}j, 
\label{eq:LatticeDynamicsEq_2}
\end{equation}
and the anharmonic Hamiltonian operator $\hat{H}_{A}$ can be written as
\begin{eqnarray} 
\hat{H}_{A} =&& \sum_{\lambda} V\left(\lambda\right) \hat{A}_{\lambda} +\sum_{n=3} \sum_{\lambda_{1}} \cdots \sum_{\lambda_{n}} \nonumber \\
&&\times V\left(\lambda_{1};\ldots;\lambda_{n}\right) \hat{A}_{\lambda_{1}}	\cdots \hat{A}_{\lambda_{n}}, \quad \lambda_{i} \equiv \vec{q}_{i}j_{i}. 
\label{eq:LatticeDynamicsEq_3} 
\end{eqnarray}
In Eq. \ref{eq:LatticeDynamicsEq_3} 
\begin{equation} 
\hat{A}_{\lambda} =   \hat{a}_{\lambda} + \hat{a}^{\dagger}_{-\lambda}, \quad -\lambda \equiv -\vec{q}j,
\label{eq:LatticeDynamicsEq_4}
\end{equation}
and \cite{Born-Huang-DynamicalTheory-1954}
\begin{eqnarray} 
V&&\left(\lambda_{1};\ldots;\lambda_{n}\right) \nonumber \\
&&= \frac{1}{n! N^{n}} \left(\frac{\hbar}{2 }\right)^{n/2} \frac{ \Delta\left( \vec{q}_{1} +\cdots+\vec{q}_{n} \right) }{ \left[\omega_{\lambda_{1}} \cdots \omega_{\lambda_{n}} \right]^{1/2} } \nonumber \\
&&\times \sum_{\kappa_{1},\alpha_{1}} \sum_{l_{2},\kappa_{2},\alpha_{2}} \cdots \sum_{l_{n},\kappa_{n},\alpha_{n}} \Phi_{\alpha_{1} \cdots \alpha_{n}}\left(0 \kappa_{1};l_{2} \kappa_{2};\ldots;l'_{n} \kappa'_{n} \right) \nonumber \\
&&\times \frac{e_{\alpha_{1}}\left(\kappa_{1}|\lambda_{1}\right)}{M^{1/2}_{\kappa_{1}}} \cdots \frac{e_{\alpha_{n}}\left(\kappa_{n}|\lambda_{n}\right)}{M^{1/2}_{\kappa_{n}}} e^{i \left[\vec{q}_{2}\cdot \vec{x}\left(l_{2}\right)+\cdots+\vec{q}_{n}\cdot \vec{x}\left(l_{n}\right) \right]}.
\label{eq:LatticeDynamicsEq_5} 
\end{eqnarray}
In Eq. \ref{eq:LatticeDynamicsEq_5} 
\begin{eqnarray} 
\Phi&&_{\alpha_{1} \cdots \alpha_{n}}\left(l_{1} \kappa_{1};\ldots;l_{n} \kappa_{n} \right) \nonumber \\
&&\equiv \left.\frac{\partial^{n}{\Phi}}{\partial{x'_{\alpha_{1}}\left(l_{1} \kappa_{1} \right)} \cdots \partial{x'_{\alpha_{n}}\left(l_{n} \kappa_{n} \right)}  } \right|_{\left\{x'\left(l_{i}\kappa_{i}\right) = x\left(l_{i}\kappa_{i}\right)\right\}},
\label{eq:LatticeDynamicsEq_6}
\end{eqnarray}
are the so-called $n$th order atomic force constants or interatomic force constants (IFCs), which are derivatives of the potential energy $\Phi$, $\left\{\alpha_{i}\right\}$ are Cartesian indices, $\left\{\vec{q}_{i}\right\}$ phonon wave vectors (the wave vector times $2 \pi$), $\left\{j_{i}\right\}$ phonon mode indices, $\left\{\vec{e}\left(\kappa_{i}|\lambda_{i}\right)\right\}$ phonon eigenvectors, $\left\{\omega_{\lambda_{i}}\right\}$ phonon eigenvalues, $\left\{M_{\kappa_{i}}\right\}$ atomic masses of atoms $\left\{\kappa_{i}\right\}$ and $\vec{x}\left(l\kappa\right) = \vec{x}\left(l\right) + \vec{x}\left(\kappa\right)$, where $\vec{x}\left(l\right)$ is the lattice translational vector and $\vec{x}\left(\kappa\right)$ the position vector of atom $\kappa$ within the unit cell. Furthermore, $\hat{a}^{\dagger}_{\lambda}$ and $\hat{a}_{\lambda}$ are the so-called creation and annihilation operators for phonons, respectively.

The diagonalization of the Hamiltonian was obtained with the following expansions for displacement and momentum
\begin{equation} 
\hat{u}_{\alpha}\left(l \kappa\right) = \left(\frac{\hbar}{2 N^{2} M_{\kappa}}\right)^{1/2} \sum_{\lambda} \omega^{-1/2}_{\lambda} e^{i \vec{q}\cdot \vec{x}\left(l\right)} e_{\alpha}\left(\kappa| \lambda\right) \hat{A}_{\lambda},
\label{eq:LatticeDynamicsEq_8}
\end{equation}
\begin{equation} 
\hat{p}_{\alpha}\left(l \kappa\right) = -i \left(\frac{\hbar M_{\kappa}}{2 N^{2} }\right)^{1/2} \sum_{\lambda} \omega^{1/2}_{\lambda} e^{i \vec{q}\cdot \vec{x}\left(l\right)} e_{\alpha}\left(\kappa| \lambda\right)  \hat{B}_{\lambda},
\label{eq:LatticeDynamicsEq_9}
\end{equation}
where $N$ is the number of $\vec{q}$-points in the $\vec{q}$-mesh and 
\begin{equation} 
\hat{B}_{\lambda} = \hat{a}_{\lambda} - \hat{a}^{\dagger}_{-\lambda}.
\label{eq:LatticeDynamicsEq_10}
\end{equation}
The phonon eigenvectors and eigenvalues can be obtained from the eigenvalue equation
\begin{equation}
\omega^{2}_{j}\left(\vec{q}\right)e_{\alpha}\left(\kappa|\vec{q}j\right) = \sum_{\kappa',\beta}D_{\alpha\beta}\left(\kappa\kappa'|\vec{q}\right)e_{\beta}\left(\kappa'|\vec{q}j\right),
\label{eq:LatticeDynamicsEq_11}   
\end{equation}
with
\begin{equation}
D_{\alpha\beta}\left(\kappa\kappa'|\vec{q}\right) \equiv \sum_{l} \frac{\Phi_{\alpha\beta}\left(l\kappa;0\kappa'\right)}{\sqrt{M_{\kappa}M_{\kappa'}}} e^{-i\vec{q}\cdot \vec{x}\left(l\right)}.
\label{eq:LatticeDynamicsEq_12}
\end{equation}
The components of the eigenvector $\vec{e}\left(\kappa|\vec{q}j\right)$ are usually chosen to satisfy the following orthonormality and closure conditions
\begin{equation}
\sum_{\kappa,\alpha}e_{\alpha}\left(\kappa|\vec{q}j'\right)e^{*}_{\alpha}\left(\kappa|\vec{q}j\right)=\delta_{jj'},
\label{eq:LatticeDynamicsEq_13}
\end{equation}
\begin{equation}
\sum_{j}e_{\alpha}\left(\kappa|\vec{q}j\right)e^{*}_{\beta}\left(\kappa'|\vec{q}j\right)=\delta_{\alpha\beta}\delta_{\kappa\kappa'},
\label{eq:LatticeDynamicsEq_14}
\end{equation}
where $\delta_{\alpha\beta}$ is the Kronecker delta. One may interpret $\vec{e}\left(\kappa|\vec{q}j\right) e^{i\vec{q}\cdot \vec{x}\left(l\right)}$ as the probability amplitude and 
\begin{equation}
\left|\vec{e}\left(\kappa|\vec{q}j\right) e^{i\vec{q}\cdot \vec{x}\left(l\right)} \right|^{2} = \left|\vec{e}\left(\kappa|\vec{q}j\right)\right|^{2},
\label{eq:LatticeDynamicsEq_15}
\end{equation}
as the probability that the atom $l\kappa$ vibrates in the phonon mode $\vec{q}j$ \cite{Harkonen-ElasticManyBodyPert-2016}.

\subsection{Thermal conductivity}
\label{Thermalconductivity}

By using the BTE approach \cite{Ziman-ElectronsPhonons-1960,Srivastava-PhysicsOfPhonons-1990}, one may write for the lattice thermal conductivity \cite{Omini-iterative-BTE-1995,Omini-PhysRevB.53.9064-iterative-BTE-1996,Ward-PhysRevB.80.125203-Tcond-2009,Li-shengbte-2014}
\begin{equation}
\kappa_{\alpha \beta} = \frac{ \hbar }{k_{B}T V} \sum_{\lambda}   \omega_{\lambda} v_{\alpha}\left(\lambda\right) \bar{n}_{\lambda} \left(\bar{n}_{\lambda} + 1\right)  F_{\beta,\lambda},
\label{eq:ThermalconductivityEq_1}
\end{equation}
where $V$ is the volume of the unit cell, $k_{B}$ is the Boltzmann constant, $\vec{v}\left(\lambda\right)$ is the phonon group velocity and $\bar{n}_{\lambda}$ the equilibrium distribution function for the state $\lambda$. The unknown term  $F_{\beta,\lambda}$ is obtained by solving the iterative equation 
\begin{eqnarray}
F_{\alpha,\lambda} =&& \frac{1}{X_{\lambda}} \sum_{\lambda'} \sum_{\lambda''} \left[  \Gamma^{\lambda''}_{\lambda \lambda'} \left(F_{\alpha,\lambda''} -F_{\alpha,\lambda'} \right) \right. \nonumber \\
&&+ \left. \Gamma^{\lambda' \lambda''}_{\lambda }  \left(F_{\alpha,\lambda'}+F_{\alpha,\lambda''} \right) \right] \nonumber \\
&&+ \frac{ \hbar \omega_{\lambda} v_{\alpha}\left(\lambda\right) }{ T X_{\lambda} } \bar{n}_{\lambda} \left(\bar{n}_{\lambda} + 1\right), 
\label{eq:ThermalconductivityEq_2}
\end{eqnarray}  
where (when only three phonon scattering is included)
\begin{equation}
X_{\lambda} \equiv \sum_{\lambda'} \sum_{\lambda''} \left( \Gamma^{\lambda''}_{\lambda \lambda'} + \Gamma^{\lambda' \lambda''}_{\lambda }  \right).
\label{eq:ThermalconductivityEq_3}
\end{equation}  
In Eqs. \ref{eq:ThermalconductivityEq_2} and \ref{eq:ThermalconductivityEq_3}, $\Gamma^{\lambda' \lambda''}_{\lambda }$  is the transition probability for processes in which a phonon $\lambda$ vanishes and two phonons $\lambda',\lambda''$ are created. Accordingly, $\Gamma^{\lambda''}_{\lambda \lambda'}$ is the transition probability for the opposite process. The transition probabilities $\Gamma^{\lambda''}_{\lambda \lambda'},\Gamma^{\lambda' \lambda''}_{\lambda }$ can be obtained, for instance, from the golden rule or from the phonon self energy \cite{Maradudin-Fein-ScatteringOfNeutrons-1962,Li-shengbte-2014}. For example, one may write (here $\beta = 1/k_{B} T$)
\begin{eqnarray} 
\sum_{\lambda'}  \sum_{\lambda''} \Gamma^{\lambda' \lambda''}_{\lambda }  =&&   18 \frac{\beta \pi }{\hbar} \sum_{\lambda'}  \sum_{\lambda''}  \left| V\left(\lambda;\lambda';\lambda''\right) \right|^{2}  \nonumber \\
&&\times  \left( \bar{n}_{\lambda'}+ \bar{n}_{\lambda''} +1 \right) \delta\left[ \omega_{\lambda} - \omega_{\lambda'} - \omega_{\lambda''} \right], \nonumber \\
\label{eq:ThermalconductivityEq_5_1}
\end{eqnarray}
and in a similar way for the transition rates $\Gamma^{\lambda''}_{\lambda \lambda'}$ \cite{Maradudin-Fein-ScatteringOfNeutrons-1962,Li-shengbte-2014}. In Eq. \ref{eq:ThermalconductivityEq_5_1}, the coefficients $\left| V\left(\lambda;\lambda';\lambda''\right) \right|^{2}$ are given by Eq. \ref{eq:LatticeDynamicsEq_5}. By Eq. \ref{eq:LatticeDynamicsEq_5}, the decrease of the mass of the atoms $\kappa,\kappa',\kappa''$ relative to the third-order IFCs $\Phi_{\alpha \alpha' \alpha''}\left(0 \kappa;l' \kappa';\ldots;l'' \kappa'' \right)$ results in larger transition rates and shorter relaxation times (RT) in general. Also, the decrease of the phonon eigenvalues $\omega_{\lambda}, \omega_{\lambda'}, \omega_{\lambda''}$ has the same effect when other factors are fixed. These effects within the studied structures are considered in Sec. \ref{ResultsForLatticeThermalConductivityAndRelatedQuantities}. The quantity $F_{\alpha,\lambda}$ is related to the RT as \cite{Ward-PhysRevB.80.125203-Tcond-2009}
\begin{equation}
\tau_{\alpha}\left(\lambda\right) = \frac{T F_{\alpha,\lambda}}{ \hbar \omega_{\lambda} v_{\alpha}\left(\lambda\right)},
\label{eq:ThermalconductivityEq_5}
\end{equation}
thus Eq. \ref{eq:ThermalconductivityEq_1} can be written as
\begin{equation}
\kappa_{\alpha \beta} = \frac{1}{V} \sum_{\lambda}  v_{\alpha}\left(\lambda\right) v_{\beta}\left(\lambda\right) c_{v}\left(\lambda\right) \tau_{\beta}\left(\lambda\right),
\label{eq:ThermalconductivityEq_6}
\end{equation}
where the heat capacity at constant volume for the phonon state $\lambda$ may be written as
\begin{equation}
c_{v}\left(\lambda\right) = k_{B} \beta^{2} \hbar^{2}\omega^{2}_{\lambda} \bar{n}_{\lambda} \left(\bar{n}_{\lambda} + 1\right).
\label{eq:ThermalconductivityEq_7}
\end{equation}

The shortcomings of the method used to calculate the lattice thermal conductivity in this work are discussed in Ref. \cite{Harkonen-Tcond-II-VIII-PhysRevB.93.024307-2016}. Recently, computational studies for real materials, where the temperature dependence of the IFCs is taken into account have been carried out \cite{Paulatto-PhysRevB.91.054304_lifetimes-2015,Romero-TcondPbTe-PhysRevB.91.214310-2015}. This effect is neglected in the present approach and it may have some significance on the present results.

\subsection{Studied structures and computational details}
\label{StudiedStructuresAndComputationalDetails}
\begin{figure}
\includegraphics[width=0.49\textwidth]{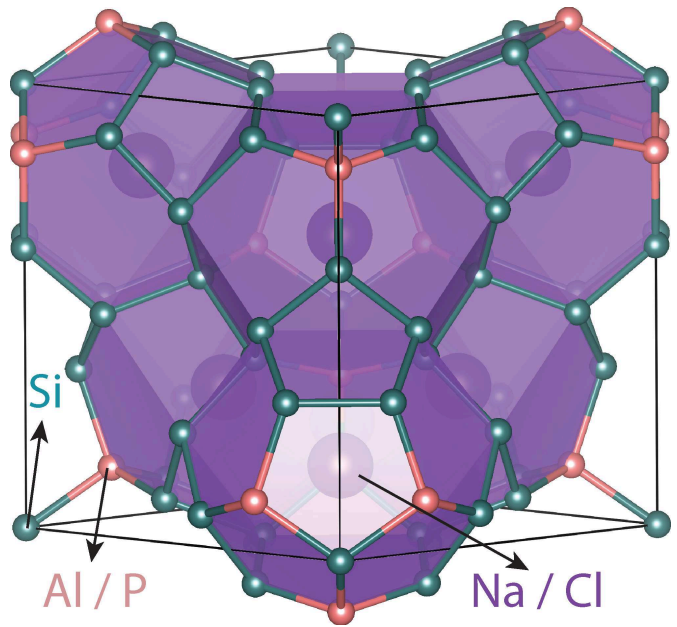}
\caption{The crystallographic body-centered cubic unit cell of the Zintl clathrates [Si$_{19}$P$_{4}$]Cl$_{4}$ and Na$_{4}$[Al$_{4}$Si$_{19}$]. The figure was prepared using the VESTA visualization program \cite{Momma-Vesta-2011}. (Color online)}
\label{fig:StructureFig}
\end{figure}

The space group of all the studied structures is $I\bar{4}3m\left(217\right)$. The silicon clathrate framework Si$_{23}$ has 23 atoms in the primitive unit cell, while the Zintl clathrates [Si$_{19}$P$_{4}$]Cl$_{4}$ and Na$_{4}$[Al$_{4}$Si$_{19}$] have 27 atoms in the primitive unit cell. The crystallographic body-centered cubic unit cell of [Si$_{19}$P$_{4}$]Cl$_{4}$ and Na$_{4}$[Al$_{4}$Si$_{19}$] with 54 atoms is illustrated in Fig. \ref{fig:StructureFig}.
The parent silicon framework can be considered to be composed of fused polyhedral cages (cavities), where the vertices correspond to four-coordinated silicon atoms. The framework heteroatoms Al/P occupy the 8c Wyckoff position within the Si framework, while the Na/Cl guest atoms are located inside the polyhedral cages (Wyckoff position 8c) \cite{Karttunen-Structuralprinc-2010}. The Na$_{4}$[Al$_{4}$Si$_{19}$] and [Si$_{19}$P$_{4}$]Cl$_{4}$ structures can be classified as anionic and cationic clathrates, respectively \cite{Shevelkov-zintl-clath-2011}. In the so-called anionic Zintl clathrates such as Na4[Al4Si19], there is a charge transfer from the less electronegative guest atoms (Na) to the framework atoms (Si-Al). In the so-called cationic Zintl clathrates such as [Si19P4]Cl4, there is a charge transfer from the less electronegative  framework atoms (Si-P) to the guest atoms (Cl). The bonding within the framework can be considered to be covalent, while the framework-guest interactions are of ionic nature \cite{Gatti-GuestFrameworkInt-2003}.

The \textit{ab initio} density functional calculations to optimize the crystal structures and to calculate the phonon eigenvalues and eigenvectors were carried out with the Quantum Espresso program package (QE, version 5.0.3).\cite{QE-2009} Atoms were described using ultrasoft pseudopotentials and plane wave basis set\cite{Garrity-pseudopotentials-2014}. The Generalized Gradient Approximation (GGA) was applied by using the PBE exchange-correlation energy functionals \cite{Perdew-generalized-1996}. If not otherwise mentioned, the applied computational parameters and methods were similar to those used in Ref. \cite{Harkonen-Tcond-II-VIII-PhysRevB.93.024307-2016}. The results for the clathrate framework Si$_{23}$, used here for comparative analysis of the Zintl clathrates, were taken from Ref. \cite{Harkonen-Tcond-II-VIII-PhysRevB.93.024307-2016}. A (6,6,6) mesh was used for the electronic $\vec{k}$ sampling, while the $\vec{q}$ meshes used phonon and lattice thermal conductivity calculations were (4,4,4) and (10,10,10), respectively. Both the lattice constants and the atomic positions of the studied structures were optimized by forcing the space group $I\bar{4}3m$. The optimized lattice constants were 10.10 \AA, 10.00 \AA, and 10.37 \AA~for Si$_{23}$, [Si$_{19}$P$_{4}$]Cl$_{4}$, and Na$_{4}$[Al$_{4}$Si$_{19}$], respectively. The non-analytic corrections to dynamical matrices in the limit $\vec{q} \rightarrow 0$ were taken into account in the QE and ShengBTE calculations. The version 1.0.2 of ShengBTE was used. In the lattice thermal conductivity calculations, three-phonon and isotopic scattering were included. The constant scalebroad was set to 0.5 in all ShengBTE calculations \cite{Li-Gaussian_PhysRevB.85.195436-2012,Li-shengbte-2014}. For all structures, third-order IFCs were calculated up to 6th-nearest neigbours using the program thirdorder.py included in the ShengBTE distribution \cite{Li-ThirdOrderPy_PhysRevB.86.174307-2012}. A $\left(3,3,3\right)$ supercell was used to calculate the third-order IFCs with thirdorder.py in all cases.

The validity of the present computational approach was assessed in Ref. \cite{Harkonen-Tcond-II-VIII-PhysRevB.93.024307-2016}, for instance, by comparing the calculated lattice thermal conductivity values to the experimental ones in the case of the silicon diamond structure (\textit{d}-Si). The difference between the calculated and experimental values was about 4\%-13\% within the temperature range 125-300 K, the smallest differences being obtained at higher temperatures.

\section{Results and discussion}
\label{cha:ResultsAndDiscussion}

\subsection{Phonon spectrum}
\label{PhononSpectrum}
\begin{figure}
\includegraphics[width=0.49\textwidth]{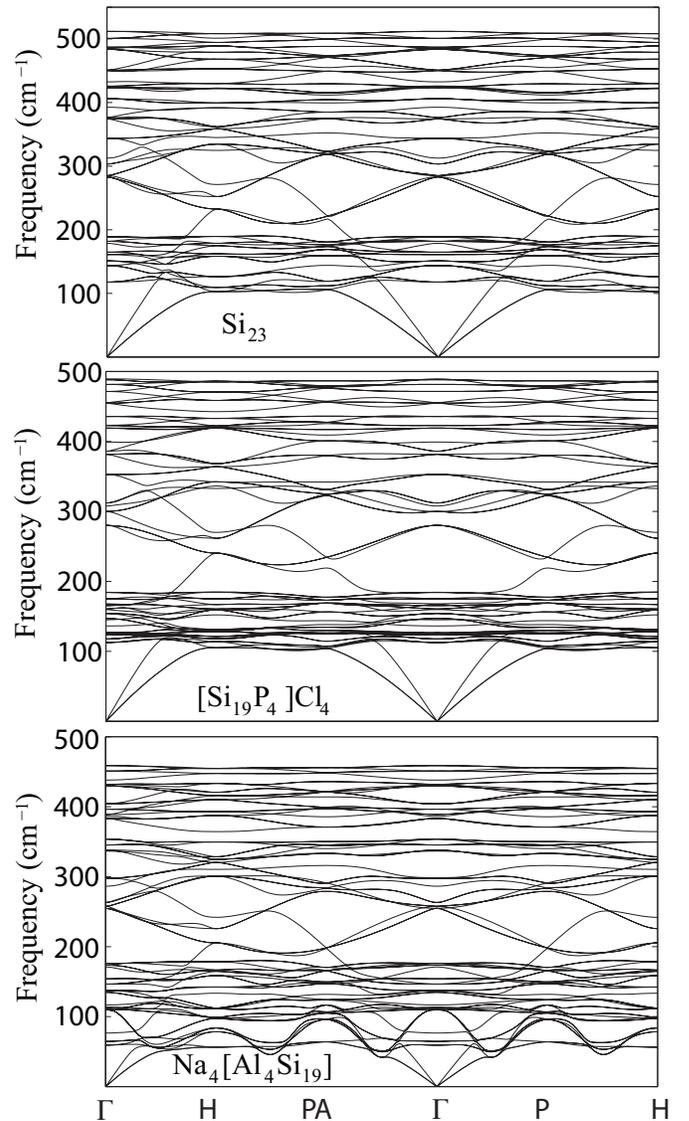}
\caption{Phonon eigenvalues as a function of wave vector (dispersion relations) along high symmetry paths in the first Brillouin zone for Si$_{23}$, [Si$_{19}$P$_{4}$]Cl$_{4}$ and Na$_{4}$[Al$_{4}$Si$_{19}$].}
\label{fig:DispersionFig}
\end{figure}

The calculated phonon eigenvalues (dispersion relations) along high symmetry paths for the structures Si$_{23}$, [Si$_{19}$P$_{4}$]Cl$_{4}$ and Na$_{4}$[Al$_{4}$Si$_{19}$] are shown in Fig. \ref{fig:DispersionFig}. 
The phonon dispersions are rather similar for all structures within the frequency range 200-500 cm$^{-1}$. For frequencies below 100 cm$^{-1}$, Si$_{23}$ and [Si$_{19}$P$_{4}$]Cl$_{4}$ show rather similar spectrum while the spectrum for Na$_{4}$[Al$_{4}$Si$_{19}$] appears to be different. The differences between $Si_{23}$ and [Si$_{19}$P$_{4}$]Cl$_{4}$ are mostly due to the Cl guest atoms within the frequency range 100-200 cm$^{-1}$ (this can be seen from the atom projected phonon density of states considered later in this section). The maximum frequencies of acoustic modes for Na$_{4}$[Al$_{4}$Si$_{19}$] are about two times smaller than the corresponding values in the case of Si$_{23}$ and [Si$_{19}$P$_{4}$]Cl$_{4}$. Furthermore, the acoustic and lowest-energy optical modes of the structure Na$_{4}$[Al$_{4}$Si$_{19}$] show oscillatory behaviour, for example, along the high symmetry paths $\Gamma-P$ and $\Gamma-PA$, while in the case of Si$_{23}$ and [Si$_{19}$P$_{4}$]Cl$_{4}$ this behaviour seems to be absent.

According to Eqs. \ref{eq:LatticeDynamicsEq_11} and \ref{eq:LatticeDynamicsEq_13} and since both structures with guest atoms have the same exponential factors, $e^{-i\vec{q}\cdot \vec{x}\left(l\right)}$, difference of the dynamical matrix elements $\left\{D_{\alpha\beta}\left(\kappa\kappa'|\vec{q}\right)\right\}$ is due to the second-order IFCs divided by the atomic masses $D_{\alpha\beta}\left(l\kappa;0\kappa'\right) = \Phi_{\alpha\beta}\left(l\kappa;0\kappa'\right)/\sqrt{M_{\kappa}M_{\kappa'}}$ (Eq. \ref{eq:LatticeDynamicsEq_12}). After one fixes the dynamical matrix, the eigenvectors can be calculated numerically: this and the preceding in mind one may infer that the differences in the dispersion relations for frequencies below 100 cm$^{-1}$ indicate differences in $\left\{D_{\alpha\beta}\left(l\kappa;0\kappa'\right)\right\}$ between the considered structures.

\begin{figure}
\includegraphics[width=0.49\textwidth]{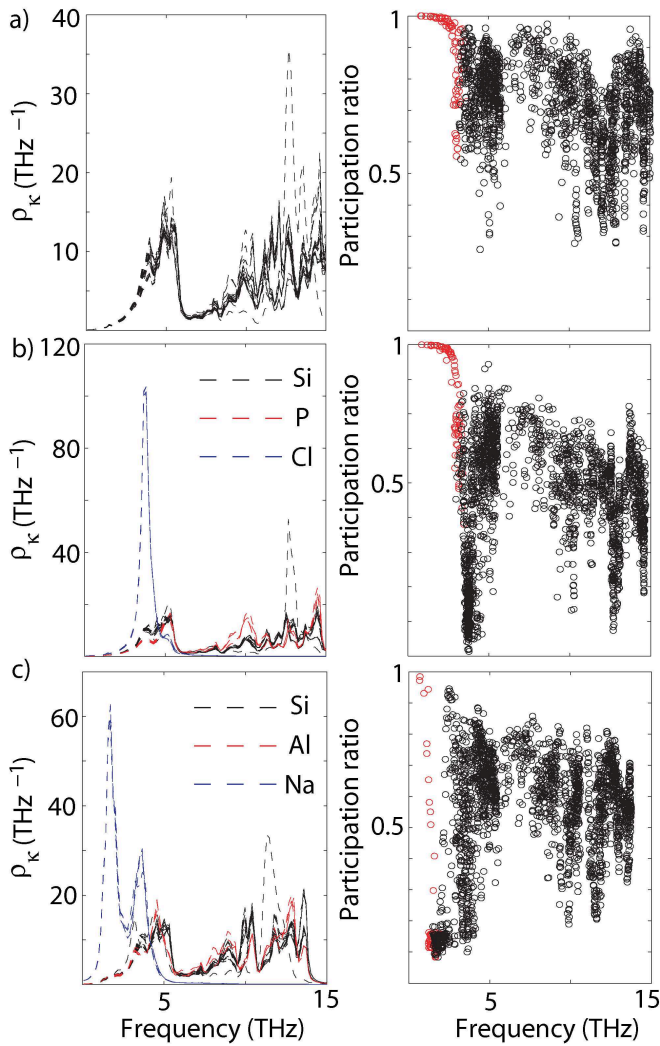}
\caption{The atom projected density of states $\rho_{\kappa}\left(\omega\right)$ and participation ratio for each structure. (a) Si$_{23}$, (b) [Si$_{19}$P$_{4}$]Cl$_{4}$ and (c) Na$_{4}$[Al$_{4}$Si$_{19}$]. The participation ratio values for the acoustic modes are drawn in red. (Color online)}
\label{fig:PrPDOSParticipationAllStructuresFigure}
\end{figure}

In Fig. \ref{fig:PrPDOSParticipationAllStructuresFigure}, the so-called atom projected density of states $\rho_{\kappa}\left(\omega\right) = 1/N \sum_{\lambda} \left|\vec{e}\left(\kappa|\lambda\right)\right|^{2} \delta\left(\omega-\omega_{\lambda}\right)$ and the participation ratio (PR) defined as \cite{Bell-Localization-0022-3719-3-10-013-1970,Hafner-PropagatingExQuasiCryst-1993,Pailhes-Localization-PhysRevLett.113.025506-2014,Tadano-ImpactOfRattlers-PhysRevLett.114.095501-2015}
\begin{equation}
PR \, = \frac{\left[\sum_{\kappa} \left|\vec{e}\left(\kappa|\lambda\right)  \right|^{2} M^{-1}_{\kappa} \right]^{2}}{N_{a} \sum_{\kappa} \left|\vec{e}\left(\kappa|\lambda\right)  \right|^{4} M^{-2}_{\kappa} },
\label{eq:PhononSpectrumEq_2}   
\end{equation}
are shown to further analyze the differences in the phonon spectrum of the studied structures. 
$N_{a}$ is the number of atoms within the unit cell. If one considers $\left|\vec{e}\left(\kappa|\lambda\right)\right|^{2}$ as the probability distribution (Sec. \ref{LatticeDynamics}), then $\rho_{\kappa}\left(\omega\right)$ may be considered as the expected value of the phonon density of states (PDOS) for each $\kappa$. It can be seen from Fig. \ref{fig:PrPDOSParticipationAllStructuresFigure}, that the contribution of the framework heteroatoms (Al and P) to $\rho_{\kappa}\left(\omega\right)$ is rather similar to that of the Si framework atoms, when comparing all three structures. However, some differences can be identified, for instance, at the frequencies 3-6 THz. The guest atoms Cl and Na mainly contribute to the phonon modes at the frequencies below 5 THz. Further, for Na$_{4}$[Al$_{4}$Si$_{19}$], there is a relatively large contribution of Na guest atoms at 2 THz, while an analogous contribution from the Cl guest atoms is absent in the case of [Si$_{19}$P$_{4}$]Cl$_{4}$.

The PR can be used to study the localization of the phonon modes \cite{Bell-Localization-0022-3719-3-10-013-1970}. The modes with rather local characteristics are expected to have the PR values near $N^{-1}_{a}$ (only few atoms are displaced in the mode), while the PR values of about 1 indicate the opposite. In Si$_{23}$ and [Si$_{19}$P$_{4}$]Cl$_{4}$, the PR values for the acoustic modes are approximately between 1 and 0.5, while for Na$_{4}$[Al$_{4}$Si$_{19}$] PR values as low as 0.1 are obtained and most of the PR values for the acoustic modes are clustered between 0.1 and 0.2. This indicates that the acoustic modes of Na$_{4}$[Al$_{4}$Si$_{19}$] seem to have more local characteristics than the acoustic modes in the other two structures.

To estimate the difference in the harmonic interactions, the following mean values for the second-order IFCs are used  
\begin{equation}
D\left(\kappa\right) \equiv \frac{1}{9 N_{a}}\sum_{\alpha,\beta,\kappa'} \left|D_{\alpha \beta}\left(0\kappa;0\kappa'\right)\right|.
\label{eq:PhononSpectrumEq_3}   
\end{equation}
The values $D\left(\kappa\right)$ on average for the Na guest atoms are about 25\% smaller than those for the Cl guest atoms. Also, the Al framework heteroatoms atoms show approximately 21\% smaller $D\left(\kappa\right)$ values on average than obtained for the P heteroatoms. Further, the Si framework atoms have about 10\% smaller $D\left(\kappa\right)$ values on average in the structures [Si$_{19}$P$_{4}$]Cl$_{4}$ and Na$_{4}$[Al$_{4}$Si$_{19}$] in comparison to the framework heteroatoms P and Al. Lastly, the Na and Cl guest atoms have approximately five times smaller $D\left(\kappa\right)$ values on average than obtained for the framework atoms, that is, the ratio is almost the same in both cases.

To summarize, the framework heteroatoms and the Si framework atoms seem to have rather similar effect on the phonon spectrum in all the studied structures. Furthermore, the Na and Cl guest atoms appear to have a rather large effect on the phonon spectrum at the frequencies below 5 THz and Na guest atoms seem to flatten the acoustic phonon dispersion relations in a more distinct manner in comparison to the Cl guest atoms.

\subsection{Results for lattice thermal conductivity and related quantities}
\label{ResultsForLatticeThermalConductivityAndRelatedQuantities}
\begin{figure}
\includegraphics[width=0.49\textwidth]{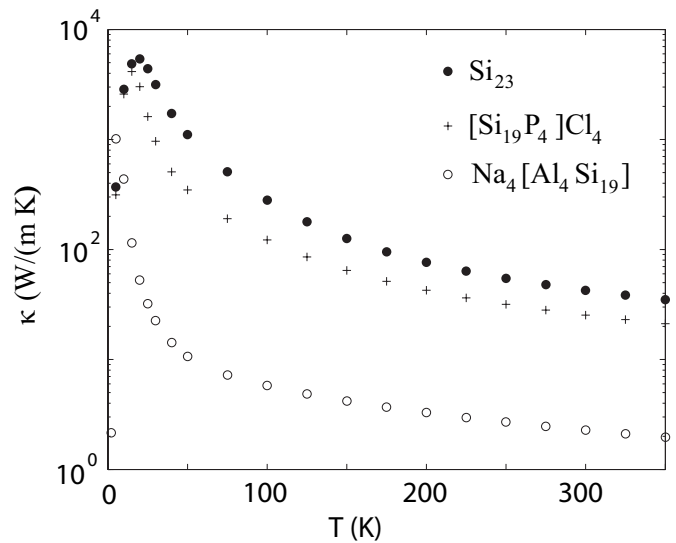}
\caption{The calculated lattice thermal conductivity values for the structures Si$_{23}$, [Si$_{19}$P$_{4}$]Cl$_{4}$ and Na$_{4}$[Al$_{4}$Si$_{19}$].}
\label{fig:KappaFig}
\end{figure}

The calculated thermal conductivity values as a function of temperature for all studied structures are shown in Fig. \ref{fig:KappaFig}.
The lattice thermal conductivities for Si$_{23}$ and [Si$_{19}$P$_{4}$]Cl$_{4}$ within the temperature range 100-300 K are 282-43 W/(m K) and  122-25 W/(m K), respectively. Within the same temperature range, the lattice thermal conductivities for Na$_{4}$[Al$_{4}$Si$_{19}$] are approximately 6-2 W/(m K). Thus, the present results indicate that Na$_{4}$[Al$_{4}$Si$_{19}$] has about twenty times smaller lattice thermal conductivity values in comparison to Si$_{23}$ and about ten times smaller values in comparison to [Si$_{19}$P$_{4}$]Cl$_{4}$. In Refs. \cite{Norouzzadeh-CTEpaper-2016,Madsen-LatCondQHA-2016}, about three times smaller lattice thermal conductivity at 300 K were obtained for Si$_{23}$ (or Si$_{46}$-VIII) by using more approximate models for the lattice conductivity based on the linearized BTE than used in the present work. Further, about four times lower lattice thermal conductivity at 300 K were obtained for the clathrates Ba$_{8}$Al$_{16}$Si$_{30}$ and Ba$_{8}$Cu$_{6}$Si$_{40}$ in Ref. \cite{Madsen-LatCondQHA-2016}, than obtained here for Na$_{4}$[Al$_{4}$Si$_{19}$].

\begin{figure*}
\includegraphics[width=0.99\textwidth]{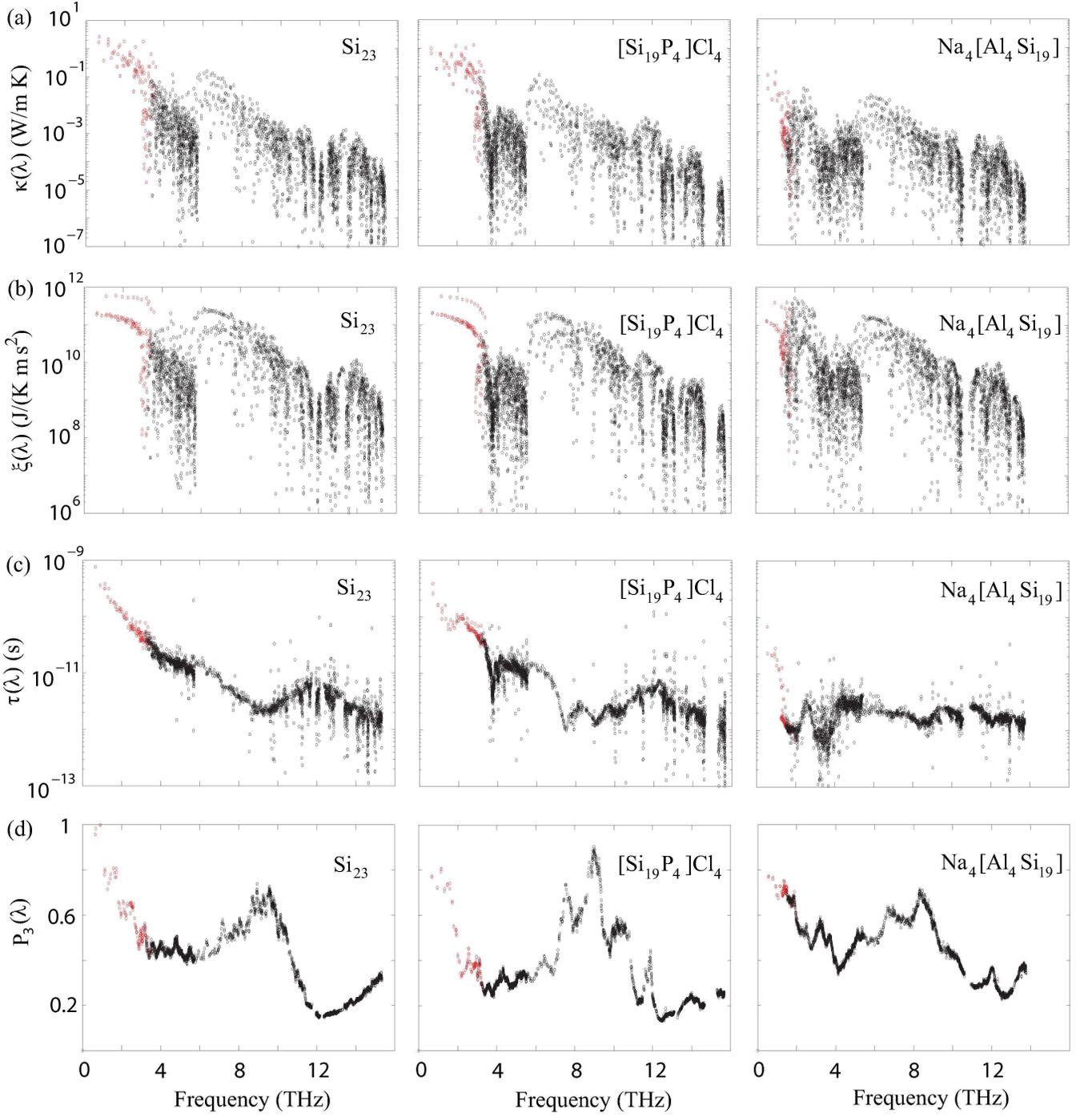}
\caption{The calculated values for each state $\lambda$ as a function of phonon frequency for the structures Si$_{23}$, [Si$_{19}$P$_{4}$]Cl$_{4}$ and Na$_{4}$[Al$_{4}$Si$_{19}$] at 300 K. (a) The lattice thermal conductivity $\kappa\left(\lambda\right) = 1/3\sum_{\alpha}\kappa_{\alpha \alpha}\left(\lambda\right)$ (negative values are not shown), (b) quantities $\xi\left(\lambda\right) \equiv 1/\left(3 V\right) \sum_{\alpha}  v^{2}_{\alpha}\left(\lambda\right) c_{v}\left(\lambda\right)$, (c) relaxation times $\tau\left(\lambda\right) \equiv  1/3 \sum_{\alpha}  \tau_{\alpha}\left(\lambda\right)$ (negative values are not shown) and (d) phonon phase space $P_{3}\left(\lambda\right)$. For all quantities, the acoustic modes are drawn in red and the optical modes in black. The reported $P_{3}\left(\lambda\right)$ values are unitless relative values obtained from $P_{3}\left(\lambda\right)/\max \left\{ \left.P_{3}\left(\lambda\right)\right|_{Si_{23}}\right\}$. (Color online)}
\label{fig:VIIIKappaTauXi}
\end{figure*}

The lattice thermal conductivity values for each state $\lambda$ in conjunction with the RTs $\tau\left(\lambda\right)$, phonon phase space $P_{3}\left(\lambda\right)$, and quantities $\xi\left(\lambda\right)$ are shown in Fig. \ref{fig:VIIIKappaTauXi}.
For Na$_{4}$[Al$_{4}$Si$_{19}$], the largest lattice thermal conductivity contributions of acoustic modes are rather systematically smaller than in the case of Si$_{23}$ or [Si$_{19}$P$_{4}$]Cl$_{4}$. A similar difference can be seen between the structures Si$_{23}$ and [Si$_{19}$P$_{4}$]Cl$_{4}$. Si$_{23}$ and [Si$_{19}$P$_{4}$]Cl$_{4}$ show rather similar values for the quantity $\xi\left(\lambda\right)$. As in the case of lattice thermal conductivity, Na$_{4}$[Al$_{4}$Si$_{19}$] also has a rather different different distribution of values of the quantity $\xi\left(\lambda\right)$, the values for acoustic modes being mostly smaller in comparison to Si$_{23}$ or [Si$_{19}$P$_{4}$]Cl$_{4}$. This is not that surprising as such because the harmonic phonon spectrum for the acoustic modes of Na$_{4}$[Al$_{4}$Si$_{19}$] is rather different in comparison to the other two structures (Fig. \ref{fig:DispersionFig}). The flattening of the acoustic modes as a function of $\vec{q}$ has ($\omega\left(\lambda\right)$ have smaller values for acoustic modes), for example, the following effects on $\xi\left(\lambda\right)$ at fixed temperature $T_{0}$: smaller values of $\omega\left(\lambda\right)$ increase $\bar{n}_{\lambda}$ and thus $c_{v}\left(\lambda\right)$. The flattening decreases the group velocity $\vec{v}\left(\lambda\right)$, thus the flattening has opposite effect on $\vec{v}\left(\lambda\right)$ and $c_{v}\left(\lambda\right)$. In the case of Na$_{4}$[Al$_{4}$Si$_{19}$], the change in the harmonic phonon spectrum seems to favor the reduced group velocities more than the increase of $c_{v}\left(\lambda\right)$, resulting in the smaller values of $\xi\left(\lambda\right)$ for acoustic modes.

The RTs, shown in Fig. \ref{fig:VIIIKappaTauXi}, reveal some differences between Si$_{23}$ and [Si$_{19}$P$_{4}$]Cl$_{4}$, the former having larger maximum values of $\tau\left(\lambda\right)$, in particular for acoustic modes. This seems to be the main reason for the different values of lattice thermal conductivity obtained for [Si$_{19}$P$_{4}$]Cl$_{4}$ and Si$_{23}$. The RTs for Na$_{4}$[Al$_{4}$Si$_{19}$] are in general smaller than obtained for [Si$_{19}$P$_{4}$]Cl$_{4}$ and Si$_{23}$. Compared with [Si$_{19}$P$_{4}$]Cl$_{4}$, the maximum values for the acoustic modes of Na$_{4}$[Al$_{4}$Si$_{19}$] are about ten times smaller. For [Si$_{19}$P$_{4}$]Cl$_{4}$ and Na$_{4}$[Al$_{4}$Si$_{19}$], it appears that the third-order coefficients $V\left(\lambda;\lambda';\lambda''\right)$ may have larger values than in the case of Si$_{23}$.

The distribution of phonon phase space values $P_{3}\left(\lambda\right) \propto \tau^{-1}\left(\lambda\right)$, also shown in Fig. \ref{fig:VIIIKappaTauXi}, are rather different for all structures despite the rather similar phonon spectra for the structures Si$_{23}$ and [Si$_{19}$P$_{4}$]Cl$_{4}$. The maximum values are, perhaps surprisingly, largest for Si$_{23}$. For example, Si$_{23}$ has larger $P_{3}\left(\lambda\right)$ values for acoustic modes than \textit{d}-Si \cite{Harkonen-Tcond-II-VIII-PhysRevB.93.024307-2016}, which probably is one of the reasons behing the smaller lattice thermal conductivity of Si$_{23}$ in comparison to \textit{d}-Si. The maximum values of $P_{3}\left(\lambda\right)$ are rather similar for [Si$_{19}$P$_{4}$]Cl$_{4}$ and Na$_{4}$[Al$_{4}$Si$_{19}$]. However, for Na$_{4}$[Al$_{4}$Si$_{19}$], the P$_{3}$ values for acoustic modes are more clustered than for the other structures and there are practically no values below 0.6, while for other structures there are rather many states with values smaller than this. Thus, it seems that the P$_{3}$ values are one the reasons behind the smaller RTs obtained for Na$_{4}$[Al$_{4}$Si$_{19}$].

\begin{figure}
\includegraphics[width=0.49\textwidth]{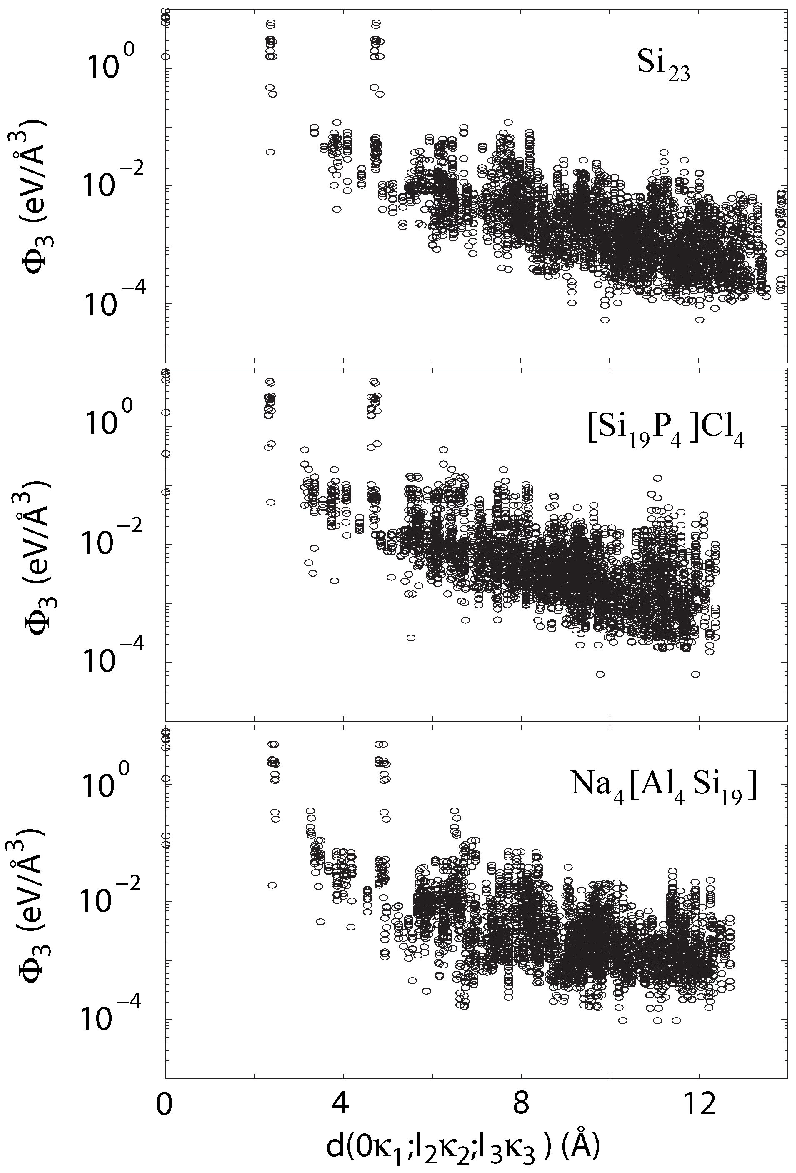}
\caption{The calculated third-order IFCs for Si$_{23}$, [Si$_{19}$P$_{4}$]Cl$_{4}$ and Na$_{4}$[Al$_{4}$Si$_{19}$]. Here, $d\left(0\kappa_{1};l_{2}\kappa_{2};l_{3}\kappa_{3}\right)$ and $\Phi_{3}$ are given by Eqs. \ref{eq:3rdOrdIFCsEq_1} and \ref{eq:3rdOrdIFCsEq_2}.}
\label{fig:3rdOrdIFCsFig}
\end{figure}

In Fig. \ref{fig:3rdOrdIFCsFig}, the third-order IFCs as a function of distance are shown.
The following quantities are used in Fig. \ref{fig:3rdOrdIFCsFig}
\begin{eqnarray} 
d&&\left(0\kappa_{1};l_{2}\kappa_{2};l_{3}\kappa_{3}\right) \nonumber \\
 &&\equiv \left|x\left(\kappa_{1}\right) - x\left(l_{2}\kappa_{2}\right)\right| + \left|x\left(\kappa_{1}\right) - x\left(l_{3}\kappa_{3}\right)\right|,
\label{eq:3rdOrdIFCsEq_1}
\end{eqnarray}
\begin{equation} 
\Phi_{3} \equiv \frac{1}{27} \sum_{\alpha_{1},\alpha_{2},\alpha_{3}} \left|\Phi_{\alpha_{1}\alpha_{2}\alpha_{3}}\left(0\kappa_{1};l_{2}\kappa_{2};l_{3}\kappa_{3}\right)\right|.
\label{eq:3rdOrdIFCsEq_2}
\end{equation}
As can be seen, the third-order IFCs of the three studied structures do not show such large differences that could have been expected based on the differences in the calculated RTs. Therefore, it seems that the different RTs are mostly due to the harmonic quantities included in the anharmonic Hamiltonian and some possible reasons for the different RTs of Na$_{4}$[Al$_{4}$Si$_{19}$] are discussed next. As mentioned in Sec. \ref{Thermalconductivity}, the decrease of mass of the atoms and the term $\omega_{\lambda}\omega_{\lambda'}\omega_{\lambda'}$ in general decreases the value of the RTs. Also, as $\omega_{\lambda'}$ and $\omega_{\lambda''}$ have smaller values, the following term $\left( \bar{n}_{\lambda'}+ \bar{n}_{\lambda''} +1 \right)$ in Eq. \ref{eq:ThermalconductivityEq_5_1} further decreases the value of the RTs. Since the PR values for the acoustic modes in the case of Na$_{4}$[Al$_{4}$Si$_{19}$] are rather small (only few atoms vibrate in a particular state), the probabilities $\left|\vec{e}\left(\kappa|\lambda\right)\right|^{2}$ and thus the phonon eigenvectors $\vec{e}\left(\kappa|\lambda\right)$ for the acoustic modes are expected to be rather large, which in turn decreases the RTs of these modes through the coefficients $\left|V\left(\lambda;\lambda';\lambda''\right)\right|^{2}$. The exponential factors, $e^{i \vec{q}_{i} \cdot \vec{x}\left(l_{i}\right)}$, are identical in all structures. Together with the differing P$_{3}$ and smaller group velocity values of the acoustic modes, these mentioned factors may have some significance in explaining the smaller RT and lattice thermal conductivity values of Na$_{4}$[Al$_{4}$Si$_{19}$]. To summarize, the stronger anharmonicity of the structure Na$_{4}$[Al$_{4}$Si$_{19}$] in comparison to Si$_{23}$ and [Si$_{19}$P$_{4}$]Cl$_{4}$ seems to arise mostly from the differing harmonic quantities instead of the third-order IFCs.

One way to measure the anharmonicity of a structure are the so-called Gr\"{u}neisen parameters. By using the perturbation theory, it has been shown that the Gr\"{u}neisen parameters can be written as \cite{Barron-DynamicalPropertiesOfSolids-1974}
\begin{eqnarray} 
\gamma_{\mu \nu}\left(\lambda\right) =  &&-\sum^{3n}_{j'=4} \frac{12 V_{\mu \nu}\left(0j'\right) V\left(0j';\lambda;-\lambda\right) }{ \hbar^{2} \omega_{0j'} \omega_{\lambda} } \nonumber \\
                                        &&- \frac{2 V_{\mu \nu} \left(\lambda;-\lambda\right) }{\hbar \omega_{\lambda}},
\label{eq:GruneisenParametersEq_1}
\end{eqnarray}
where the first term on the right hand side vanishes if the position of every atom in the unit cell is determined by the symmetry (no internal strain). In Eq. \ref{eq:GruneisenParametersEq_1}, the coefficients like $V\left(\lambda;\lambda';\lambda''\right)$ are given by Eq. \ref{eq:LatticeDynamicsEq_5} and 
\begin{eqnarray} 
V_{\mu \nu} \left(\lambda;\lambda'\right) =&& \frac{\hbar}{4 } \sum_{\kappa,\alpha} \sum_{l', \kappa',\alpha'} \sum_{l'', \kappa''} \Phi_{\alpha \alpha' \mu}\left(0 \kappa;l' \kappa'; l'' \kappa'' \right) \nonumber \\
 &&\times    \frac{e_{\alpha}\left(\kappa| \lambda \right) e_{\alpha'}\left(\kappa'| \lambda' \right) }{ \sqrt{ M_{\kappa} M_{\kappa'}  \omega_{\lambda} \omega_{\lambda'} } }  e^{i \vec{q}' \cdot \vec{x}\left(l'\right)}  x_{\nu}\left(l''\kappa''\right).  \nonumber \\
\label{eq:GruneisenParametersEq_2}
\end{eqnarray}
The Gr\"{u}neisen parameters can also be written as
\begin{equation} 
\gamma_{\mu \nu}\left(\lambda\right) = -\frac{1}{\omega_{\lambda}} \frac{\partial{\omega_{\lambda}}}{\partial{ \eta_{\mu \nu} }},
\label{eq:GruneisenParametersEq_3}
\end{equation}
and furthermore, in the case of cubic crystals
\begin{equation} 
\frac{1}{3}\gamma_{\mu \mu}\left(\lambda\right) = -\frac{V}{\omega_{\lambda}} \frac{\partial{\omega_{\lambda}}}{\partial{V}} \equiv \gamma\left(\lambda\right).
\label{eq:GruneisenParametersEq_4}
\end{equation}
When the phonon-phonon interaction is approximated in the so-called continuum theory, it has been shown that the square of the (averaged) Gr\"{u}neisen parameter is inversely proportional to the mean free path and thus the lifetime of phonons \cite{Ziman-ElectronsPhonons-1960}.

The Gr\"{u}neisen parameters are sometimes used for calculating the thermal expansion of materials. There is some evidence that, for instance, crystalline materials which have negative thermal expansion (NTE) over rather wide temperature ranges can have relatively low lattice thermal conductivity values \cite{Kennedy-NTE-thermal-cond-ZrW2O8-2005,Kennedy-NTE-thermal-cond-HfMo2O8-2007} (measured for polycrystalline samples, however). This behaviour of NTE materials would be rather logical since, as mentioned, in the continuum theory $\tau \propto \gamma^{-2}$ ($\tau$ and $\gamma$ are some average values) and within the so-called quasi-harmonic approximation (QHA), the coefficient of thermal expansion (CTE) can be written as (see for example Refs. \cite{Wallace-ThermodynamicsOfCrystals-1972}) $\alpha_{\mu_{1}\nu_{1}} = \sum_{\mu_{2},\nu_{2}}  \sum_{\lambda} s^{T}_{\mu_{1}\nu_{1} \mu_{2}\nu_{2}} c_{v}\left(\lambda\right)  \gamma_{\mu_{1}\nu_{1}}\left(\lambda\right)$, where $s^{T}_{\mu_{1}\nu_{1} \mu_{2}\nu_{2}}$ is the isothermal second-order compliance tensor, inverse to the isothermal second-order elastic constant $c^{T}_{\mu_{1}\nu_{1} \mu_{2}\nu_{2}}$. That is, the $\gamma_{\mu\nu}\left(\lambda\right)$ are usually expected to have relatively large absolute values for materials which have rather large absolute value of the CTE. There is some evidence that the silicon clathrate framework VII possesses rather unusual NTE behavior, while Si$_{23}$ (or VIII), for instance, has CTE that is rather similar to \textit{d}-Si \cite{Harkonen-NTE-2014}.

\begin{figure}
\includegraphics[width=0.49\textwidth]{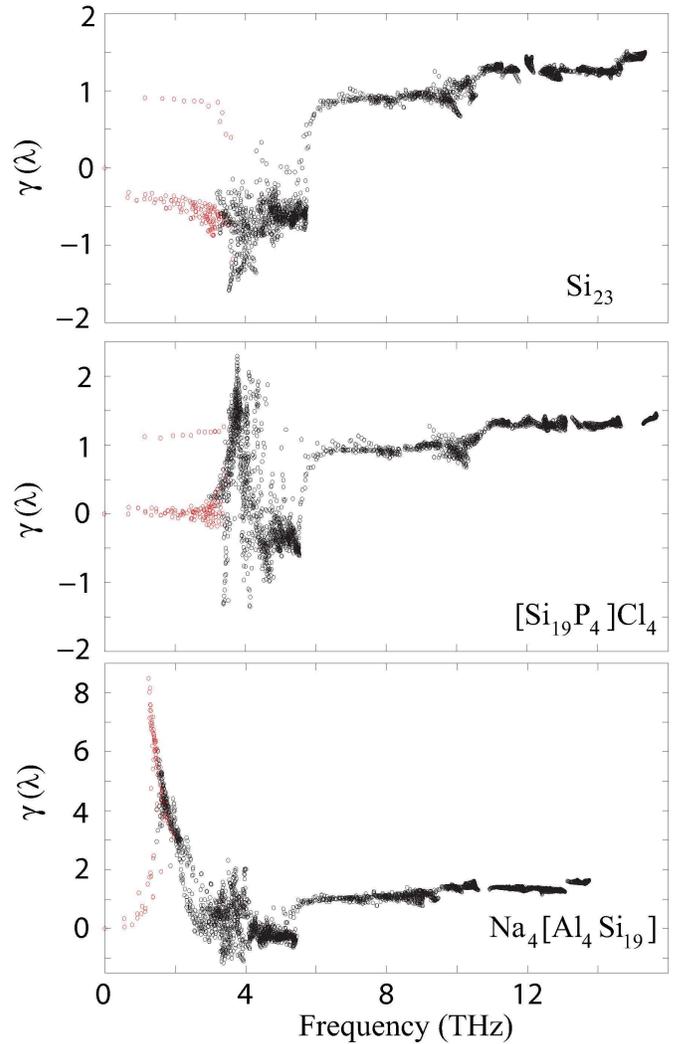}
\caption{The calculated Gr\"{u}neisen parameter values for Si$_{23}$, [Si$_{19}$P$_{4}$]Cl$_{4}$ and Na$_{4}$[Al$_{4}$Si$_{19}$]. The acoustic modes are shown in red and the optical modes in black. (Color online)}
\label{fig:GruneisenFig}
\end{figure}

The calculated Gr\"{u}neisen parameter values $\gamma\left(\lambda\right)$ as a function of frequency for each structure are shown in Fig. \ref{fig:GruneisenFig}.
For acoustic modes, the distribution of $\gamma\left(\lambda\right)$ values in the case of Si$_{23}$ and [Si$_{19}$P$_{4}$]Cl$_{4}$ is rather similar, while for Na$_{4}$[Al$_{4}$Si$_{19}$] fairly different result is obtained. Compared with Na$_{4}$[Al$_{4}$Si$_{19}$], the maximum values of $\gamma\left(\lambda\right)$ are about eight times smaller in the case of Si$_{23}$ and [Si$_{19}$P$_{4}$]Cl$_{4}$. For Na$_{4}$[Al$_{4}$Si$_{19}$], the lowest-energy optical modes have about three times larger Gr\"{u}neisen parameter values in comparison to [Si$_{19}$P$_{4}$]Cl$_{4}$. For Si$_{23}$, the Gr\"{u}neisen parameter results are similar to those obtained in Refs. \cite{Harkonen-NTE-2014} and \cite{Norouzzadeh-CTEpaper-2016}.

\begin{figure}
\includegraphics[width=0.49\textwidth]{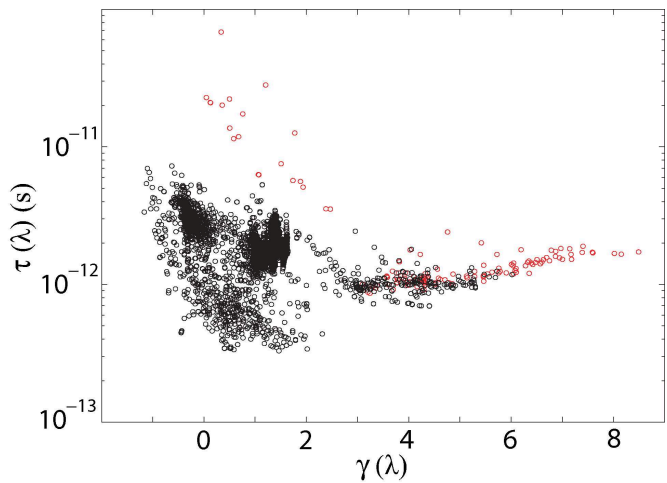}
\caption{The calculated RTs and the corresponding Gr\"{u}neisen parameter values for each state $\lambda$ in Na$_{4}$[Al$_{4}$Si$_{19}$]. The acoustic modes are shown in red and the optical modes in black. (Color online)}
\label{fig:TauGrunNa4Si19Al4Fig}
\end{figure}

To study the relationship between the RTs and Gr\"{u}neisen parameters, these quantities are depicted in Fig. \ref{fig:TauGrunNa4Si19Al4Fig} for Na$_{4}$[Al$_{4}$Si$_{19}$]. 
For the acoustic modes, when $\gamma\left(\lambda\right) \geq 2$, the maximum $\tau\left(\lambda\right)$ values are approximately $2 \times 10^{-12}$ s, which is about one order of magnitude smaller than the largest values obtained. Thus, these modes have a rather small contribution to the lattice thermal conductivity. The smallest $\tau\left(\lambda\right)$ values for the acoustic modes are obtained when the Gr\"{u}neisen parameters have values between 3 and 4. The relationship shown in Fig. \ref{fig:TauGrunNa4Si19Al4Fig} indicates that in some cases, there can be a connection between the relatively large absolute values of CTE and rather low lattice thermal conductivities.

As already mentioned in Sec. \ref{cha:Introduction}, several mechanisms have been proposed to explain the rather small lattice thermal conductivity of various Zintl clathrates. In Ref. \cite{Christensen-AvoidedCrossing-2008}, for instance, it was concluded for the Ba$_8$Ga$_{16}$Ge$_{30}$ clathrate based on inelastic neutron scattering data that the reduction of the lattice thermal conductivity is mostly due to the flattening of the phonon dispersion relations caused by the guest atoms instead the shorter RTs of the phonons. In Ref. \cite{Pailhes-Localization-PhysRevLett.113.025506-2014}, inelastic x-ray scattering and \textit{ab initio} lattice dynamical studies on the Ba$_8$Si$_{46}$ clathrate resulted in the conclusion that the reduction of the lattice thermal conductivity follows from the changes in the harmonic spectrum induced by the guest-framework interactions and that the reduced RTs have a rather small significance. In contrast, in an \textit{ab initio} lattice dynamical study for the Ba$_8$Ga$_{16}$Ge$_{30}$ clathrate \cite{Tadano-ImpactOfRattlers-PhysRevLett.114.095501-2015}, the largest reduction in the lattice thermal conductivity was suggested to arise from the smaller RT values, while the reduction of the phonon group velocities was found to have a smaller effect (the BTE was not solved iteratively, but within the single-mode relaxation time approximation).

The present results show similarities with the results obtained in Ref. \cite{Tadano-ImpactOfRattlers-PhysRevLett.114.095501-2015} for the Ba$_8$Ga$_{16}$Ge$_{30}$ clathrate and in Ref. \cite{Li-UltraLowLatCond-PhysRevB.91.144304-2015} for the YbFe$_{4}$Sb$_{12}$ skutterudite. In the case of YbFe$_{4}$Sb$_{12}$,\cite{Li-UltraLowLatCond-PhysRevB.91.144304-2015} it was summarized that the increased phonon scattering is due to the differing phonon phase space, third-order IFCs having rather marginal effect on the reduction of the lattice thermal conductivity.  The results of the present work show that for materials with relatively similar third-order IFCs, there can be rather significant changes in anharmonicity and in the lattice thermal conductivity values, which essentially follow from the differing second-order IFCs. To sum up some central findings of the present work: in the case of Na$_{4}$[Al$_{4}$Si$_{19}$], the second-order IFCs seem to produce the harmonic phonon spectrum such that three-phonon phase space favors the phonon scattering and the phonon eigenvectors $\vec{e}\left(\kappa|\lambda\right)$ have rather large values for phonons of smallest frequencies (localization). These effects facilitate the reduction of RTs and phonon group velocities, which in turn leads to the reduced lattice thermal conductivity values. The hypothetical structures studied here possess rather high symmetry (symmetry is forced in the calculation) and it is probable that possible structural disorder decreases the lattice thermal conductivity even further \cite{Chakoumakos-StructuralDisordLatCond-2000,Chakoumakos-StructuralDisordClath-2001}.

\section{Conclusions}
\label{cha:Conclusions}
The lattice thermal conductivity of the silicon clathrate framework Si$_{23}$ and two Zintl clathrates, [Si$_{19}$P$_{4}$]Cl$_{4}$ and Na$_{4}$[Al$_{4}$Si$_{19}$], was investigated by using \textit{ab initio} lattice dynamics together with an iterative solution of the linearized BTE. The lattice thermal conductivity of the structure Na$_{4}$[Al$_{4}$Si$_{19}$] was found to be about one order of magnitude lower at 300 K in comparison to the other two materials studied here. The lower lattice thermal conductivity of Na$_{4}$[Al$_{4}$Si$_{19}$]  is mostly due to lower relaxation times and phonon group velocities, which differ from Si$_{23}$ and [Si$_{19}$P$_{4}$]Cl$_{4}$ largely due second-order IFCs. Furthermore, it appears that the anharmonicity of two similar crystalline materials can be rather different from one another mostly because of differing second-order IFCs. The present results may shed light on the understanding about the lattice thermal conductivity of clathrates and skutterudites, for example, which can give further guidance for the discovery of more efficient thermoelectric materials.

\begin{acknowledgments}
We gratefully acknowledge funding from the Foundation for Research of Natural Resources in Finland (grant 17591/13). The computing resources for this work were provided by the Finnish Grid Infrastructure (FGI) and CSC - the Finnish IT Center for Science. 
\end{acknowledgments}
\bibliography{bibfile}

\begin{thebibliography}{62}%
\makeatletter
\providecommand \@ifxundefined [1]{%
 \@ifx{#1\undefined}
}%
\providecommand \@ifnum [1]{%
 \ifnum #1\expandafter \@firstoftwo
 \else \expandafter \@secondoftwo
 \fi
}%
\providecommand \@ifx [1]{%
 \ifx #1\expandafter \@firstoftwo
 \else \expandafter \@secondoftwo
 \fi
}%
\providecommand \natexlab [1]{#1}%
\providecommand \enquote  [1]{``#1''}%
\providecommand \bibnamefont  [1]{#1}%
\providecommand \bibfnamefont [1]{#1}%
\providecommand \citenamefont [1]{#1}%
\providecommand \href@noop [0]{\@secondoftwo}%
\providecommand \href [0]{\begingroup \@sanitize@url \@href}%
\providecommand \@href[1]{\@@startlink{#1}\@@href}%
\providecommand \@@href[1]{\endgroup#1\@@endlink}%
\providecommand \@sanitize@url [0]{\catcode `\\12\catcode `\$12\catcode
  `\&12\catcode `\#12\catcode `\^12\catcode `\_12\catcode `\%12\relax}%
\providecommand \@@startlink[1]{}%
\providecommand \@@endlink[0]{}%
\providecommand \url  [0]{\begingroup\@sanitize@url \@url }%
\providecommand \@url [1]{\endgroup\@href {#1}{\urlprefix }}%
\providecommand \urlprefix  [0]{URL }%
\providecommand \Eprint [0]{\href }%
\providecommand \doibase [0]{http://dx.doi.org/}%
\providecommand \selectlanguage [0]{\@gobble}%
\providecommand \bibinfo  [0]{\@secondoftwo}%
\providecommand \bibfield  [0]{\@secondoftwo}%
\providecommand \translation [1]{[#1]}%
\providecommand \BibitemOpen [0]{}%
\providecommand \bibitemStop [0]{}%
\providecommand \bibitemNoStop [0]{.\EOS\space}%
\providecommand \EOS [0]{\spacefactor3000\relax}%
\providecommand \BibitemShut  [1]{\csname bibitem#1\endcsname}%
\let\auto@bib@innerbib\@empty
\bibitem [{\citenamefont {Ioffe}(1957)}]{Ioffe-SemicondThermoelements-1957}%
  \BibitemOpen
  \bibfield  {author} {\bibinfo {author} {\bibfnamefont {A.~F.}\ \bibnamefont
  {Ioffe}},\ }\href@noop {} {\emph {\bibinfo {title} {Semiconductor
  Thermoelements and Thermoelectric Cooling}}}\ (\bibinfo  {publisher}
  {Infosearch Limited London},\ \bibinfo {year} {1957})\ pp.\ \bibinfo {pages}
  {1--184}\BibitemShut {NoStop}%
\bibitem [{\citenamefont {Slack}\ and\ \citenamefont
  {Rowe}(1995)}]{Slack-crtThermoelectrics-1995}%
  \BibitemOpen
  \bibfield  {author} {\bibinfo {author} {\bibfnamefont {G.~A.}\ \bibnamefont
  {Slack}}\ and\ \bibinfo {author} {\bibfnamefont {D.}~\bibnamefont {Rowe}},\
  }\href@noop {} {\bibfield  {journal} {\bibinfo  {journal} {CRC, Boca Raton,
  FL}\ ,\ \bibinfo {pages} {407}} (\bibinfo {year} {1995})}\BibitemShut
  {NoStop}%
\bibitem [{\citenamefont {Kasper}\ \emph {et~al.}(1965)\citenamefont {Kasper},
  \citenamefont {Hagenmuller}, \citenamefont {Pouchard},\ and\ \citenamefont
  {Cros}}]{Kasper-24121965.science.clath.1965}%
  \BibitemOpen
  \bibfield  {author} {\bibinfo {author} {\bibfnamefont {J.~S.}\ \bibnamefont
  {Kasper}}, \bibinfo {author} {\bibfnamefont {P.}~\bibnamefont {Hagenmuller}},
  \bibinfo {author} {\bibfnamefont {M.}~\bibnamefont {Pouchard}}, \ and\
  \bibinfo {author} {\bibfnamefont {C.}~\bibnamefont {Cros}},\ }\href {\doibase
  10.1126/science.150.3704.1713} {\bibfield  {journal} {\bibinfo  {journal}
  {Science}\ }\textbf {\bibinfo {volume} {150}},\ \bibinfo {pages} {1713}
  (\bibinfo {year} {1965})}\BibitemShut {NoStop}%
\bibitem [{\citenamefont {Nolas}\ \emph {et~al.}(1998)\citenamefont {Nolas},
  \citenamefont {Cohn}, \citenamefont {Slack},\ and\ \citenamefont
  {Schujman}}]{Nolas-Ge-clath-thermoelect-1998}%
  \BibitemOpen
  \bibfield  {author} {\bibinfo {author} {\bibfnamefont {G.}~\bibnamefont
  {Nolas}}, \bibinfo {author} {\bibfnamefont {J.}~\bibnamefont {Cohn}},
  \bibinfo {author} {\bibfnamefont {G.}~\bibnamefont {Slack}}, \ and\ \bibinfo
  {author} {\bibfnamefont {S.}~\bibnamefont {Schujman}},\ }\href {\doibase
  10.1063/1.121747} {\bibfield  {journal} {\bibinfo  {journal} {Appl. Phys.
  Lett.}\ }\textbf {\bibinfo {volume} {73}},\ \bibinfo {pages} {178} (\bibinfo
  {year} {1998})}\BibitemShut {NoStop}%
\bibitem [{\citenamefont {Kovnir}\ and\ \citenamefont
  {Shevelkov}(2004)}]{Kovnir-0036-021X-73-9-R06.clath.review-2004}%
  \BibitemOpen
  \bibfield  {author} {\bibinfo {author} {\bibfnamefont {K.~A.}\ \bibnamefont
  {Kovnir}}\ and\ \bibinfo {author} {\bibfnamefont {A.~V.}\ \bibnamefont
  {Shevelkov}},\ }\href {http://stacks.iop.org/0036-021X/73/i=9/a=R06}
  {\bibfield  {journal} {\bibinfo  {journal} {Russ. Chem. Rev.}\ }\textbf
  {\bibinfo {volume} {73}},\ \bibinfo {pages} {923} (\bibinfo {year}
  {2004})}\BibitemShut {NoStop}%
\bibitem [{\citenamefont {Karttunen}\ \emph {et~al.}(2010)\citenamefont
  {Karttunen}, \citenamefont {Fässler}, \citenamefont {Linnolahti},\ and\
  \citenamefont {Pakkanen}}]{Karttunen-Structuralprinc-2010}%
  \BibitemOpen
  \bibfield  {author} {\bibinfo {author} {\bibfnamefont {A.~J.}\ \bibnamefont
  {Karttunen}}, \bibinfo {author} {\bibfnamefont {T.~F.}\ \bibnamefont
  {Fässler}}, \bibinfo {author} {\bibfnamefont {M.}~\bibnamefont
  {Linnolahti}}, \ and\ \bibinfo {author} {\bibfnamefont {T.~A.}\ \bibnamefont
  {Pakkanen}},\ }\href@noop {} {\bibfield  {journal} {\bibinfo  {journal}
  {Inorg. Chem.}\ }\textbf {\bibinfo {volume} {50}},\ \bibinfo {pages} {1733}
  (\bibinfo {year} {2010})}\BibitemShut {NoStop}%
\bibitem [{\citenamefont {Takabatake}\ \emph {et~al.}(2014)\citenamefont
  {Takabatake}, \citenamefont {Suekuni}, \citenamefont {Nakayama},\ and\
  \citenamefont {Kaneshita}}]{Takabatake-RevModPhys.86.669-2014}%
  \BibitemOpen
  \bibfield  {author} {\bibinfo {author} {\bibfnamefont {T.}~\bibnamefont
  {Takabatake}}, \bibinfo {author} {\bibfnamefont {K.}~\bibnamefont {Suekuni}},
  \bibinfo {author} {\bibfnamefont {T.}~\bibnamefont {Nakayama}}, \ and\
  \bibinfo {author} {\bibfnamefont {E.}~\bibnamefont {Kaneshita}},\ }\href
  {\doibase 10.1103/RevModPhys.86.669} {\bibfield  {journal} {\bibinfo
  {journal} {Rev. Mod. Phys.}\ }\textbf {\bibinfo {volume} {86}},\ \bibinfo
  {pages} {669} (\bibinfo {year} {2014})}\BibitemShut {NoStop}%
\bibitem [{\citenamefont {Norouzzadeh}\ \emph {et~al.}(2014)\citenamefont
  {Norouzzadeh}, \citenamefont {Myles},\ and\ \citenamefont
  {Vashaee}}]{Norouzzadeh-GiantPowerFactorClatVIII-2014}%
  \BibitemOpen
  \bibfield  {author} {\bibinfo {author} {\bibfnamefont {P.}~\bibnamefont
  {Norouzzadeh}}, \bibinfo {author} {\bibfnamefont {C.~W.}\ \bibnamefont
  {Myles}}, \ and\ \bibinfo {author} {\bibfnamefont {D.}~\bibnamefont
  {Vashaee}},\ }\href@noop {} {\bibfield  {journal} {\bibinfo  {journal} {Sci.
  Rep.}\ }\textbf {\bibinfo {volume} {4}},\ \bibinfo {pages} {7028} (\bibinfo
  {year} {2014})}\BibitemShut {NoStop}%
\bibitem [{\citenamefont {Cohn}\ \emph {et~al.}(1999)\citenamefont {Cohn},
  \citenamefont {Nolas}, \citenamefont {Fessatidis}, \citenamefont {Metcalf},\
  and\ \citenamefont {Slack}}]{Cohn-GlasslikeHeatCond-PhysRevLett.82.779-1999}%
  \BibitemOpen
  \bibfield  {author} {\bibinfo {author} {\bibfnamefont {J.~L.}\ \bibnamefont
  {Cohn}}, \bibinfo {author} {\bibfnamefont {G.~S.}\ \bibnamefont {Nolas}},
  \bibinfo {author} {\bibfnamefont {V.}~\bibnamefont {Fessatidis}}, \bibinfo
  {author} {\bibfnamefont {T.~H.}\ \bibnamefont {Metcalf}}, \ and\ \bibinfo
  {author} {\bibfnamefont {G.~A.}\ \bibnamefont {Slack}},\ }\href {\doibase
  10.1103/PhysRevLett.82.779} {\bibfield  {journal} {\bibinfo  {journal} {Phys.
  Rev. Lett.}\ }\textbf {\bibinfo {volume} {82}},\ \bibinfo {pages} {779}
  (\bibinfo {year} {1999})}\BibitemShut {NoStop}%
\bibitem [{\citenamefont {Tse}\ \emph {et~al.}(2000)\citenamefont {Tse},
  \citenamefont {Uehara}, \citenamefont {Rousseau}, \citenamefont {Ker},
  \citenamefont {Ratcliffe}, \citenamefont {White},\ and\ \citenamefont
  {MacKay}}]{Tse-TcondClath-PhysRevLett.85.114-2000}%
  \BibitemOpen
  \bibfield  {author} {\bibinfo {author} {\bibfnamefont {J.~S.}\ \bibnamefont
  {Tse}}, \bibinfo {author} {\bibfnamefont {K.}~\bibnamefont {Uehara}},
  \bibinfo {author} {\bibfnamefont {R.}~\bibnamefont {Rousseau}}, \bibinfo
  {author} {\bibfnamefont {A.}~\bibnamefont {Ker}}, \bibinfo {author}
  {\bibfnamefont {C.~I.}\ \bibnamefont {Ratcliffe}}, \bibinfo {author}
  {\bibfnamefont {M.~A.}\ \bibnamefont {White}}, \ and\ \bibinfo {author}
  {\bibfnamefont {G.}~\bibnamefont {MacKay}},\ }\href {\doibase
  10.1103/PhysRevLett.85.114} {\bibfield  {journal} {\bibinfo  {journal} {Phys.
  Rev. Lett.}\ }\textbf {\bibinfo {volume} {85}},\ \bibinfo {pages} {114}
  (\bibinfo {year} {2000})}\BibitemShut {NoStop}%
\bibitem [{\citenamefont {Dong}\ \emph {et~al.}(2001)\citenamefont {Dong},
  \citenamefont {Sankey},\ and\ \citenamefont
  {Myles}}]{Dong-LatCondGeClath-PhysRevLett.86.2361-2001}%
  \BibitemOpen
  \bibfield  {author} {\bibinfo {author} {\bibfnamefont {J.}~\bibnamefont
  {Dong}}, \bibinfo {author} {\bibfnamefont {O.~F.}\ \bibnamefont {Sankey}}, \
  and\ \bibinfo {author} {\bibfnamefont {C.~W.}\ \bibnamefont {Myles}},\ }\href
  {\doibase 10.1103/PhysRevLett.86.2361} {\bibfield  {journal} {\bibinfo
  {journal} {Phys. Rev. Lett.}\ }\textbf {\bibinfo {volume} {86}},\ \bibinfo
  {pages} {2361} (\bibinfo {year} {2001})}\BibitemShut {NoStop}%
\bibitem [{\citenamefont {Nolas}\ \emph {et~al.}(2003)\citenamefont {Nolas},
  \citenamefont {Beekman}, \citenamefont {Gryko}, \citenamefont {Lamberton~Jr},
  \citenamefont {Tritt},\ and\ \citenamefont
  {McMillan}}]{Nolas-ThermalCondSi136-2003}%
  \BibitemOpen
  \bibfield  {author} {\bibinfo {author} {\bibfnamefont {G.}~\bibnamefont
  {Nolas}}, \bibinfo {author} {\bibfnamefont {M.}~\bibnamefont {Beekman}},
  \bibinfo {author} {\bibfnamefont {J.}~\bibnamefont {Gryko}}, \bibinfo
  {author} {\bibfnamefont {G.}~\bibnamefont {Lamberton~Jr}}, \bibinfo {author}
  {\bibfnamefont {T.}~\bibnamefont {Tritt}}, \ and\ \bibinfo {author}
  {\bibfnamefont {P.}~\bibnamefont {McMillan}},\ }\href@noop {} {\bibfield
  {journal} {\bibinfo  {journal} {Appl. Phys. lett.}\ }\textbf {\bibinfo
  {volume} {82}},\ \bibinfo {pages} {910} (\bibinfo {year} {2003})}\BibitemShut
  {NoStop}%
\bibitem [{\citenamefont {Bentien}\ \emph {et~al.}(2004)\citenamefont
  {Bentien}, \citenamefont {Christensen}, \citenamefont {Bryan}, \citenamefont
  {Sanchez}, \citenamefont {Paschen}, \citenamefont {Steglich}, \citenamefont
  {Stucky},\ and\ \citenamefont
  {Iversen}}]{Bentien-TcondClath-PhysRevB.69.045107-2004}%
  \BibitemOpen
  \bibfield  {author} {\bibinfo {author} {\bibfnamefont {A.}~\bibnamefont
  {Bentien}}, \bibinfo {author} {\bibfnamefont {M.}~\bibnamefont
  {Christensen}}, \bibinfo {author} {\bibfnamefont {J.~D.}\ \bibnamefont
  {Bryan}}, \bibinfo {author} {\bibfnamefont {A.}~\bibnamefont {Sanchez}},
  \bibinfo {author} {\bibfnamefont {S.}~\bibnamefont {Paschen}}, \bibinfo
  {author} {\bibfnamefont {F.}~\bibnamefont {Steglich}}, \bibinfo {author}
  {\bibfnamefont {G.~D.}\ \bibnamefont {Stucky}}, \ and\ \bibinfo {author}
  {\bibfnamefont {B.~B.}\ \bibnamefont {Iversen}},\ }\href {\doibase
  10.1103/PhysRevB.69.045107} {\bibfield  {journal} {\bibinfo  {journal} {Phys.
  Rev. B}\ }\textbf {\bibinfo {volume} {69}},\ \bibinfo {pages} {045107}
  (\bibinfo {year} {2004})}\BibitemShut {NoStop}%
\bibitem [{\citenamefont {Avila}\ \emph {et~al.}(2006)\citenamefont {Avila},
  \citenamefont {Suekuni}, \citenamefont {Umeo}, \citenamefont {Fukuoka},
  \citenamefont {Yamanaka},\ and\ \citenamefont
  {Takabatake}}]{Avila-GlassLikeTCond-PhysRevB.74.125109-2006}%
  \BibitemOpen
  \bibfield  {author} {\bibinfo {author} {\bibfnamefont {M.~A.}\ \bibnamefont
  {Avila}}, \bibinfo {author} {\bibfnamefont {K.}~\bibnamefont {Suekuni}},
  \bibinfo {author} {\bibfnamefont {K.}~\bibnamefont {Umeo}}, \bibinfo {author}
  {\bibfnamefont {H.}~\bibnamefont {Fukuoka}}, \bibinfo {author} {\bibfnamefont
  {S.}~\bibnamefont {Yamanaka}}, \ and\ \bibinfo {author} {\bibfnamefont
  {T.}~\bibnamefont {Takabatake}},\ }\href {\doibase
  10.1103/PhysRevB.74.125109} {\bibfield  {journal} {\bibinfo  {journal} {Phys.
  Rev. B}\ }\textbf {\bibinfo {volume} {74}},\ \bibinfo {pages} {125109}
  (\bibinfo {year} {2006})}\BibitemShut {NoStop}%
\bibitem [{\citenamefont {Suekuni}\ \emph {et~al.}(2007)\citenamefont
  {Suekuni}, \citenamefont {Avila}, \citenamefont {Umeo},\ and\ \citenamefont
  {Takabatake}}]{Suekuni-CageSize-PhysRevB.75.195210-2007}%
  \BibitemOpen
  \bibfield  {author} {\bibinfo {author} {\bibfnamefont {K.}~\bibnamefont
  {Suekuni}}, \bibinfo {author} {\bibfnamefont {M.~A.}\ \bibnamefont {Avila}},
  \bibinfo {author} {\bibfnamefont {K.}~\bibnamefont {Umeo}}, \ and\ \bibinfo
  {author} {\bibfnamefont {T.}~\bibnamefont {Takabatake}},\ }\href {\doibase
  10.1103/PhysRevB.75.195210} {\bibfield  {journal} {\bibinfo  {journal} {Phys.
  Rev. B}\ }\textbf {\bibinfo {volume} {75}},\ \bibinfo {pages} {195210}
  (\bibinfo {year} {2007})}\BibitemShut {NoStop}%
\bibitem [{\citenamefont {Takasu}\ \emph {et~al.}(2008)\citenamefont {Takasu},
  \citenamefont {Hasegawa}, \citenamefont {Ogita}, \citenamefont {Udagawa},
  \citenamefont {Avila}, \citenamefont {Suekuni},\ and\ \citenamefont
  {Takabatake}}]{Takasu-TypeIclathTCond-PhysRevLett.100.165503-2008}%
  \BibitemOpen
  \bibfield  {author} {\bibinfo {author} {\bibfnamefont {Y.}~\bibnamefont
  {Takasu}}, \bibinfo {author} {\bibfnamefont {T.}~\bibnamefont {Hasegawa}},
  \bibinfo {author} {\bibfnamefont {N.}~\bibnamefont {Ogita}}, \bibinfo
  {author} {\bibfnamefont {M.}~\bibnamefont {Udagawa}}, \bibinfo {author}
  {\bibfnamefont {M.~A.}\ \bibnamefont {Avila}}, \bibinfo {author}
  {\bibfnamefont {K.}~\bibnamefont {Suekuni}}, \ and\ \bibinfo {author}
  {\bibfnamefont {T.}~\bibnamefont {Takabatake}},\ }\href {\doibase
  10.1103/PhysRevLett.100.165503} {\bibfield  {journal} {\bibinfo  {journal}
  {Phys. Rev. Lett.}\ }\textbf {\bibinfo {volume} {100}},\ \bibinfo {pages}
  {165503} (\bibinfo {year} {2008})}\BibitemShut {NoStop}%
\bibitem [{\citenamefont {Christensen}\ \emph {et~al.}(2008)\citenamefont
  {Christensen}, \citenamefont {Abrahamsen}, \citenamefont {Christensen},
  \citenamefont {Juranyi}, \citenamefont {Andersen}, \citenamefont {Lefmann},
  \citenamefont {Andreasson}, \citenamefont {Bahl},\ and\ \citenamefont
  {Iversen}}]{Christensen-AvoidedCrossing-2008}%
  \BibitemOpen
  \bibfield  {author} {\bibinfo {author} {\bibfnamefont {M.}~\bibnamefont
  {Christensen}}, \bibinfo {author} {\bibfnamefont {A.~B.}\ \bibnamefont
  {Abrahamsen}}, \bibinfo {author} {\bibfnamefont {N.~B.}\ \bibnamefont
  {Christensen}}, \bibinfo {author} {\bibfnamefont {F.}~\bibnamefont
  {Juranyi}}, \bibinfo {author} {\bibfnamefont {N.~H.}\ \bibnamefont
  {Andersen}}, \bibinfo {author} {\bibfnamefont {K.}~\bibnamefont {Lefmann}},
  \bibinfo {author} {\bibfnamefont {J.}~\bibnamefont {Andreasson}}, \bibinfo
  {author} {\bibfnamefont {C.~R.}\ \bibnamefont {Bahl}}, \ and\ \bibinfo
  {author} {\bibfnamefont {B.~B.}\ \bibnamefont {Iversen}},\ }\href@noop {}
  {\bibfield  {journal} {\bibinfo  {journal} {Nature Mater.}\ }\textbf
  {\bibinfo {volume} {7}},\ \bibinfo {pages} {811} (\bibinfo {year}
  {2008})}\BibitemShut {NoStop}%
\bibitem [{\citenamefont {Avila}\ \emph {et~al.}(2008)\citenamefont {Avila},
  \citenamefont {Suekuni}, \citenamefont {Umeo}, \citenamefont {Fukuoka},
  \citenamefont {Yamanaka},\ and\ \citenamefont
  {Takabatake}}]{Avila-ba8ga16sn30TCond-2008}%
  \BibitemOpen
  \bibfield  {author} {\bibinfo {author} {\bibfnamefont {M.}~\bibnamefont
  {Avila}}, \bibinfo {author} {\bibfnamefont {K.}~\bibnamefont {Suekuni}},
  \bibinfo {author} {\bibfnamefont {K.}~\bibnamefont {Umeo}}, \bibinfo {author}
  {\bibfnamefont {H.}~\bibnamefont {Fukuoka}}, \bibinfo {author} {\bibfnamefont
  {S.}~\bibnamefont {Yamanaka}}, \ and\ \bibinfo {author} {\bibfnamefont
  {T.}~\bibnamefont {Takabatake}},\ }\href@noop {} {\bibfield  {journal}
  {\bibinfo  {journal} {Appl. Phys. Lett.}\ }\textbf {\bibinfo {volume} {92}},\
  \bibinfo {pages} {041901} (\bibinfo {year} {2008})}\BibitemShut {NoStop}%
\bibitem [{\citenamefont {Christensen}\ \emph {et~al.}(2009)\citenamefont
  {Christensen}, \citenamefont {Johnsen}, \citenamefont {Juranyi},\ and\
  \citenamefont {Iversen}}]{Christensen-ClathrateGuest-2009}%
  \BibitemOpen
  \bibfield  {author} {\bibinfo {author} {\bibfnamefont {M.}~\bibnamefont
  {Christensen}}, \bibinfo {author} {\bibfnamefont {S.}~\bibnamefont
  {Johnsen}}, \bibinfo {author} {\bibfnamefont {F.}~\bibnamefont {Juranyi}}, \
  and\ \bibinfo {author} {\bibfnamefont {B.~B.}\ \bibnamefont {Iversen}},\
  }\href@noop {} {\bibfield  {journal} {\bibinfo  {journal} {J. Appl. Phys.}\
  }\textbf {\bibinfo {volume} {105}},\ \bibinfo {pages} {073508} (\bibinfo
  {year} {2009})}\BibitemShut {NoStop}%
\bibitem [{\citenamefont {Christensen}\ \emph {et~al.}(2010)\citenamefont
  {Christensen}, \citenamefont {Johnsen},\ and\ \citenamefont
  {Iversen}}]{Christensen.et.al.-2010-B916400F}%
  \BibitemOpen
  \bibfield  {author} {\bibinfo {author} {\bibfnamefont {M.}~\bibnamefont
  {Christensen}}, \bibinfo {author} {\bibfnamefont {S.}~\bibnamefont
  {Johnsen}}, \ and\ \bibinfo {author} {\bibfnamefont {B.~B.}\ \bibnamefont
  {Iversen}},\ }\href {\doibase 10.1039/B916400F} {\bibfield  {journal}
  {\bibinfo  {journal} {Dalton Trans.}\ }\textbf {\bibinfo {volume} {39}},\
  \bibinfo {pages} {978} (\bibinfo {year} {2010})}\BibitemShut {NoStop}%
\bibitem [{\citenamefont {Candolfi}\ \emph {et~al.}(2011)\citenamefont
  {Candolfi}, \citenamefont {Aydemir}, \citenamefont {Ormeci}, \citenamefont
  {Carrillo-Cabrera}, \citenamefont {Burkhardt}, \citenamefont {Baitinger},
  \citenamefont {Oeschler}, \citenamefont {Steglich},\ and\ \citenamefont
  {Grin}}]{Candolfi-TransportClathrates-2011}%
  \BibitemOpen
  \bibfield  {author} {\bibinfo {author} {\bibfnamefont {C.}~\bibnamefont
  {Candolfi}}, \bibinfo {author} {\bibfnamefont {U.}~\bibnamefont {Aydemir}},
  \bibinfo {author} {\bibfnamefont {A.}~\bibnamefont {Ormeci}}, \bibinfo
  {author} {\bibfnamefont {W.}~\bibnamefont {Carrillo-Cabrera}}, \bibinfo
  {author} {\bibfnamefont {U.}~\bibnamefont {Burkhardt}}, \bibinfo {author}
  {\bibfnamefont {M.}~\bibnamefont {Baitinger}}, \bibinfo {author}
  {\bibfnamefont {N.}~\bibnamefont {Oeschler}}, \bibinfo {author}
  {\bibfnamefont {F.}~\bibnamefont {Steglich}}, \ and\ \bibinfo {author}
  {\bibfnamefont {Y.}~\bibnamefont {Grin}},\ }\href@noop {} {\bibfield
  {journal} {\bibinfo  {journal} {J. Appl. Phys.}\ }\textbf {\bibinfo {volume}
  {110}},\ \bibinfo {pages} {043715} (\bibinfo {year} {2011})}\BibitemShut
  {NoStop}%
\bibitem [{\citenamefont {Euchner}\ \emph {et~al.}(2012)\citenamefont
  {Euchner}, \citenamefont {Pailh\`es}, \citenamefont {Nguyen}, \citenamefont
  {Assmus}, \citenamefont {Ritter}, \citenamefont {Haghighirad}, \citenamefont
  {Grin}, \citenamefont {Paschen},\ and\ \citenamefont
  {de~Boissieu}}]{Euchner-RatlingClath-PhysRevB.86.224303-2012}%
  \BibitemOpen
  \bibfield  {author} {\bibinfo {author} {\bibfnamefont {H.}~\bibnamefont
  {Euchner}}, \bibinfo {author} {\bibfnamefont {S.}~\bibnamefont {Pailh\`es}},
  \bibinfo {author} {\bibfnamefont {L.~T.~K.}\ \bibnamefont {Nguyen}}, \bibinfo
  {author} {\bibfnamefont {W.}~\bibnamefont {Assmus}}, \bibinfo {author}
  {\bibfnamefont {F.}~\bibnamefont {Ritter}}, \bibinfo {author} {\bibfnamefont
  {A.}~\bibnamefont {Haghighirad}}, \bibinfo {author} {\bibfnamefont
  {Y.}~\bibnamefont {Grin}}, \bibinfo {author} {\bibfnamefont {S.}~\bibnamefont
  {Paschen}}, \ and\ \bibinfo {author} {\bibfnamefont {M.}~\bibnamefont
  {de~Boissieu}},\ }\href {\doibase 10.1103/PhysRevB.86.224303} {\bibfield
  {journal} {\bibinfo  {journal} {Phys. Rev. B}\ }\textbf {\bibinfo {volume}
  {86}},\ \bibinfo {pages} {224303} (\bibinfo {year} {2012})}\BibitemShut
  {NoStop}%
\bibitem [{\citenamefont {Fulmer}\ \emph {et~al.}(2013)\citenamefont {Fulmer},
  \citenamefont {Lebedev}, \citenamefont {Roddatis}, \citenamefont {Kaseman},
  \citenamefont {Sen}, \citenamefont {Dolyniuk}, \citenamefont {Lee},
  \citenamefont {Olenev},\ and\ \citenamefont
  {Kovnir}}]{Fulmer-TcondClathrate-2013}%
  \BibitemOpen
  \bibfield  {author} {\bibinfo {author} {\bibfnamefont {J.}~\bibnamefont
  {Fulmer}}, \bibinfo {author} {\bibfnamefont {O.~I.}\ \bibnamefont {Lebedev}},
  \bibinfo {author} {\bibfnamefont {V.~V.}\ \bibnamefont {Roddatis}}, \bibinfo
  {author} {\bibfnamefont {D.~C.}\ \bibnamefont {Kaseman}}, \bibinfo {author}
  {\bibfnamefont {S.}~\bibnamefont {Sen}}, \bibinfo {author} {\bibfnamefont
  {J.-A.}\ \bibnamefont {Dolyniuk}}, \bibinfo {author} {\bibfnamefont
  {K.}~\bibnamefont {Lee}}, \bibinfo {author} {\bibfnamefont {A.~V.}\
  \bibnamefont {Olenev}}, \ and\ \bibinfo {author} {\bibfnamefont
  {K.}~\bibnamefont {Kovnir}},\ }\href@noop {} {\bibfield  {journal} {\bibinfo
  {journal} {J. Am. Chem. Soc.}\ }\textbf {\bibinfo {volume} {135}},\ \bibinfo
  {pages} {12313} (\bibinfo {year} {2013})}\BibitemShut {NoStop}%
\bibitem [{\citenamefont {He}\ and\ \citenamefont
  {Galli}(2014)}]{He-NanostructuredClath-2014}%
  \BibitemOpen
  \bibfield  {author} {\bibinfo {author} {\bibfnamefont {Y.}~\bibnamefont
  {He}}\ and\ \bibinfo {author} {\bibfnamefont {G.}~\bibnamefont {Galli}},\
  }\href@noop {} {\bibfield  {journal} {\bibinfo  {journal} {Nano Lett.}\
  }\textbf {\bibinfo {volume} {14}},\ \bibinfo {pages} {2920} (\bibinfo {year}
  {2014})}\BibitemShut {NoStop}%
\bibitem [{\citenamefont {Pailh\`es}\ \emph {et~al.}(2014)\citenamefont
  {Pailh\`es}, \citenamefont {Euchner}, \citenamefont {Giordano}, \citenamefont
  {Debord}, \citenamefont {Assy}, \citenamefont {Gom\`es}, \citenamefont
  {Bosak}, \citenamefont {Machon}, \citenamefont {Paschen},\ and\ \citenamefont
  {de~Boissieu}}]{Pailhes-Localization-PhysRevLett.113.025506-2014}%
  \BibitemOpen
  \bibfield  {author} {\bibinfo {author} {\bibfnamefont {S.}~\bibnamefont
  {Pailh\`es}}, \bibinfo {author} {\bibfnamefont {H.}~\bibnamefont {Euchner}},
  \bibinfo {author} {\bibfnamefont {V.~M.}\ \bibnamefont {Giordano}}, \bibinfo
  {author} {\bibfnamefont {R.}~\bibnamefont {Debord}}, \bibinfo {author}
  {\bibfnamefont {A.}~\bibnamefont {Assy}}, \bibinfo {author} {\bibfnamefont
  {S.}~\bibnamefont {Gom\`es}}, \bibinfo {author} {\bibfnamefont
  {A.}~\bibnamefont {Bosak}}, \bibinfo {author} {\bibfnamefont
  {D.}~\bibnamefont {Machon}}, \bibinfo {author} {\bibfnamefont
  {S.}~\bibnamefont {Paschen}}, \ and\ \bibinfo {author} {\bibfnamefont
  {M.}~\bibnamefont {de~Boissieu}},\ }\href {\doibase
  10.1103/PhysRevLett.113.025506} {\bibfield  {journal} {\bibinfo  {journal}
  {Phys. Rev. Lett.}\ }\textbf {\bibinfo {volume} {113}},\ \bibinfo {pages}
  {025506} (\bibinfo {year} {2014})}\BibitemShut {NoStop}%
\bibitem [{\citenamefont {Castillo}\ \emph {et~al.}(2015)\citenamefont
  {Castillo}, \citenamefont {Schnelle}, \citenamefont {Bobnar}, \citenamefont
  {Burkhardt}, \citenamefont {B{\"o}hme}, \citenamefont {Baitinger},
  \citenamefont {Schwarz},\ and\ \citenamefont
  {Grin}}]{Castillo-Clathrate-2015}%
  \BibitemOpen
  \bibfield  {author} {\bibinfo {author} {\bibfnamefont {R.}~\bibnamefont
  {Castillo}}, \bibinfo {author} {\bibfnamefont {W.}~\bibnamefont {Schnelle}},
  \bibinfo {author} {\bibfnamefont {M.}~\bibnamefont {Bobnar}}, \bibinfo
  {author} {\bibfnamefont {U.}~\bibnamefont {Burkhardt}}, \bibinfo {author}
  {\bibfnamefont {B.}~\bibnamefont {B{\"o}hme}}, \bibinfo {author}
  {\bibfnamefont {M.}~\bibnamefont {Baitinger}}, \bibinfo {author}
  {\bibfnamefont {U.}~\bibnamefont {Schwarz}}, \ and\ \bibinfo {author}
  {\bibfnamefont {Y.}~\bibnamefont {Grin}},\ }\href@noop {} {\bibfield
  {journal} {\bibinfo  {journal} {Z. Anorg. Allg. Chem.}\ }\textbf {\bibinfo
  {volume} {641}},\ \bibinfo {pages} {206} (\bibinfo {year}
  {2015})}\BibitemShut {NoStop}%
\bibitem [{\citenamefont {Tadano}\ \emph {et~al.}(2015)\citenamefont {Tadano},
  \citenamefont {Gohda},\ and\ \citenamefont
  {Tsuneyuki}}]{Tadano-ImpactOfRattlers-PhysRevLett.114.095501-2015}%
  \BibitemOpen
  \bibfield  {author} {\bibinfo {author} {\bibfnamefont {T.}~\bibnamefont
  {Tadano}}, \bibinfo {author} {\bibfnamefont {Y.}~\bibnamefont {Gohda}}, \
  and\ \bibinfo {author} {\bibfnamefont {S.}~\bibnamefont {Tsuneyuki}},\ }\href
  {\doibase 10.1103/PhysRevLett.114.095501} {\bibfield  {journal} {\bibinfo
  {journal} {Phys. Rev. Lett.}\ }\textbf {\bibinfo {volume} {114}},\ \bibinfo
  {pages} {095501} (\bibinfo {year} {2015})}\BibitemShut {NoStop}%
\bibitem [{\citenamefont {Kishimoto}\ \emph {et~al.}(2015)\citenamefont
  {Kishimoto}, \citenamefont {Koda}, \citenamefont {Akai},\ and\ \citenamefont
  {Koyanagi}}]{Kishimoto-ThermoelectricClathII-2015}%
  \BibitemOpen
  \bibfield  {author} {\bibinfo {author} {\bibfnamefont {K.}~\bibnamefont
  {Kishimoto}}, \bibinfo {author} {\bibfnamefont {S.}~\bibnamefont {Koda}},
  \bibinfo {author} {\bibfnamefont {K.}~\bibnamefont {Akai}}, \ and\ \bibinfo
  {author} {\bibfnamefont {T.}~\bibnamefont {Koyanagi}},\ }\href@noop {}
  {\bibfield  {journal} {\bibinfo  {journal} {J. Appl. Phys.}\ }\textbf
  {\bibinfo {volume} {118}},\ \bibinfo {pages} {125103} (\bibinfo {year}
  {2015})}\BibitemShut {NoStop}%
\bibitem [{\citenamefont {Norouzzadeh}\ and\ \citenamefont
  {Myles}(2016)}]{Norouzzadeh-CTEpaper-2016}%
  \BibitemOpen
  \bibfield  {author} {\bibinfo {author} {\bibfnamefont {P.}~\bibnamefont
  {Norouzzadeh}}\ and\ \bibinfo {author} {\bibfnamefont {C.~W.}\ \bibnamefont
  {Myles}},\ }\href@noop {} {\bibfield  {journal} {\bibinfo  {journal} {J.
  Mater. Sci.}\ }\textbf {\bibinfo {volume} {51}},\ \bibinfo {pages} {4538}
  (\bibinfo {year} {2016})}\BibitemShut {NoStop}%
\bibitem [{\citenamefont {Madsen}\ \emph {et~al.}(2016)\citenamefont {Madsen},
  \citenamefont {Katre},\ and\ \citenamefont {Bera}}]{Madsen-LatCondQHA-2016}%
  \BibitemOpen
  \bibfield  {author} {\bibinfo {author} {\bibfnamefont {G.~K.}\ \bibnamefont
  {Madsen}}, \bibinfo {author} {\bibfnamefont {A.}~\bibnamefont {Katre}}, \
  and\ \bibinfo {author} {\bibfnamefont {C.}~\bibnamefont {Bera}},\ }\href@noop
  {} {\bibfield  {journal} {\bibinfo  {journal} {Phys. Status Solidi A}\
  }\textbf {\bibinfo {volume} {213}},\ \bibinfo {pages} {802} (\bibinfo {year}
  {2016})}\BibitemShut {NoStop}%
\bibitem [{\citenamefont {Huang}\ and\ \citenamefont
  {Born}(1954)}]{Born-Huang-DynamicalTheory-1954}%
  \BibitemOpen
  \bibfield  {author} {\bibinfo {author} {\bibfnamefont {K.}~\bibnamefont
  {Huang}}\ and\ \bibinfo {author} {\bibfnamefont {M.}~\bibnamefont {Born}},\
  }\href@noop {} {\emph {\bibinfo {title} {Dynamical Theory of Crystal
  Lattices}}}\ (\bibinfo  {publisher} {Clarendon Press Oxford},\ \bibinfo
  {year} {1954})\ pp.\ \bibinfo {pages} {293--306}\BibitemShut {NoStop}%
\bibitem [{\citenamefont {Maradudin}\ \emph {et~al.}(1971)\citenamefont
  {Maradudin}, \citenamefont {Montroll}, \citenamefont {Weiss},\ and\
  \citenamefont {Ipatova}}]{Maradudin-harm-appr-1971}%
  \BibitemOpen
  \bibfield  {author} {\bibinfo {author} {\bibfnamefont {A.}~\bibnamefont
  {Maradudin}}, \bibinfo {author} {\bibfnamefont {E.}~\bibnamefont {Montroll}},
  \bibinfo {author} {\bibfnamefont {G.}~\bibnamefont {Weiss}}, \ and\ \bibinfo
  {author} {\bibfnamefont {I.}~\bibnamefont {Ipatova}},\ }\href@noop {} {\emph
  {\bibinfo {title} {Theory of The Lattice Dynamics in The Harmonic
  Approximation}}},\ Vol.\ \bibinfo {volume} {Supplement 3}\ (\bibinfo
  {publisher} {Academic Press},\ \bibinfo {year} {1971})\ pp.\ \bibinfo {pages}
  {6--57}\BibitemShut {NoStop}%
\bibitem [{\citenamefont {Maradudin}\ and\ \citenamefont
  {Horton}(1974)}]{Maradudin-DynamicalPropertiesOfSolids-1974}%
  \BibitemOpen
  \bibfield  {author} {\bibinfo {author} {\bibfnamefont {A.}~\bibnamefont
  {Maradudin}}\ and\ \bibinfo {author} {\bibfnamefont {G.}~\bibnamefont
  {Horton}},\ }\href@noop {} {\emph {\bibinfo {title} {Elements of The Theory
  of Lattice Dynamics}}},\ Vol.~\bibinfo {volume} {1}\ (\bibinfo  {publisher}
  {Amsterdam: North-Holland},\ \bibinfo {year} {1974})\ pp.\ \bibinfo {pages}
  {1--82}\BibitemShut {NoStop}%
\bibitem [{\citenamefont {H\"ark\"onen}\ and\ \citenamefont
  {Karttunen}(2016)}]{Harkonen-Tcond-II-VIII-PhysRevB.93.024307-2016}%
  \BibitemOpen
  \bibfield  {author} {\bibinfo {author} {\bibfnamefont {V.~J.}\ \bibnamefont
  {H\"ark\"onen}}\ and\ \bibinfo {author} {\bibfnamefont {A.~J.}\ \bibnamefont
  {Karttunen}},\ }\href {\doibase 10.1103/PhysRevB.93.024307} {\bibfield
  {journal} {\bibinfo  {journal} {Phys. Rev. B}\ }\textbf {\bibinfo {volume}
  {93}},\ \bibinfo {pages} {024307} (\bibinfo {year} {2016})}\BibitemShut
  {NoStop}%
\bibitem [{\citenamefont
  {H{\"a}rk{\"o}nen}(2016)}]{Harkonen-ElasticManyBodyPert-2016}%
  \BibitemOpen
  \bibfield  {author} {\bibinfo {author} {\bibfnamefont {V.~J.}\ \bibnamefont
  {H{\"a}rk{\"o}nen}},\ }\href@noop {} {\bibfield  {journal} {\bibinfo
  {journal} {arXiv preprint arXiv:1603.06376}\ } (\bibinfo {year}
  {2016})}\BibitemShut {NoStop}%
\bibitem [{\citenamefont {Ziman}(1960)}]{Ziman-ElectronsPhonons-1960}%
  \BibitemOpen
  \bibfield  {author} {\bibinfo {author} {\bibfnamefont {J.~M.}\ \bibnamefont
  {Ziman}},\ }\href@noop {} {\emph {\bibinfo {title} {Electrons and Phonons:
  The Theory of Transport Phenomena in Solids}}}\ (\bibinfo  {publisher}
  {Oxford University Press},\ \bibinfo {year} {1960})\ pp.\ \bibinfo {pages}
  {264--298}\BibitemShut {NoStop}%
\bibitem [{\citenamefont
  {Srivastava}(1990)}]{Srivastava-PhysicsOfPhonons-1990}%
  \BibitemOpen
  \bibfield  {author} {\bibinfo {author} {\bibfnamefont {G.~P.}\ \bibnamefont
  {Srivastava}},\ }\href@noop {} {\emph {\bibinfo {title} {The Physics of
  Phonons}}}\ (\bibinfo  {publisher} {CRC Press},\ \bibinfo {year} {1990})\ p.\
  \bibinfo {pages} {122}\BibitemShut {NoStop}%
\bibitem [{\citenamefont {Omini}\ and\ \citenamefont
  {Sparavigna}(1995)}]{Omini-iterative-BTE-1995}%
  \BibitemOpen
  \bibfield  {author} {\bibinfo {author} {\bibfnamefont {M.}~\bibnamefont
  {Omini}}\ and\ \bibinfo {author} {\bibfnamefont {A.}~\bibnamefont
  {Sparavigna}},\ }\href@noop {} {\bibfield  {journal} {\bibinfo  {journal}
  {Physica B}\ }\textbf {\bibinfo {volume} {212}},\ \bibinfo {pages} {101}
  (\bibinfo {year} {1995})}\BibitemShut {NoStop}%
\bibitem [{\citenamefont {Omini}\ and\ \citenamefont
  {Sparavigna}(1996)}]{Omini-PhysRevB.53.9064-iterative-BTE-1996}%
  \BibitemOpen
  \bibfield  {author} {\bibinfo {author} {\bibfnamefont {M.}~\bibnamefont
  {Omini}}\ and\ \bibinfo {author} {\bibfnamefont {A.}~\bibnamefont
  {Sparavigna}},\ }\href {\doibase 10.1103/PhysRevB.53.9064} {\bibfield
  {journal} {\bibinfo  {journal} {Phys. Rev. B}\ }\textbf {\bibinfo {volume}
  {53}},\ \bibinfo {pages} {9064} (\bibinfo {year} {1996})}\BibitemShut
  {NoStop}%
\bibitem [{\citenamefont {Ward}\ \emph {et~al.}(2009)\citenamefont {Ward},
  \citenamefont {Broido}, \citenamefont {Stewart},\ and\ \citenamefont
  {Deinzer}}]{Ward-PhysRevB.80.125203-Tcond-2009}%
  \BibitemOpen
  \bibfield  {author} {\bibinfo {author} {\bibfnamefont {A.}~\bibnamefont
  {Ward}}, \bibinfo {author} {\bibfnamefont {D.~A.}\ \bibnamefont {Broido}},
  \bibinfo {author} {\bibfnamefont {D.~A.}\ \bibnamefont {Stewart}}, \ and\
  \bibinfo {author} {\bibfnamefont {G.}~\bibnamefont {Deinzer}},\ }\href
  {\doibase 10.1103/PhysRevB.80.125203} {\bibfield  {journal} {\bibinfo
  {journal} {Phys. Rev. B}\ }\textbf {\bibinfo {volume} {80}},\ \bibinfo
  {pages} {125203} (\bibinfo {year} {2009})}\BibitemShut {NoStop}%
\bibitem [{\citenamefont {Li}\ \emph {et~al.}(2014)\citenamefont {Li},
  \citenamefont {Carrete}, \citenamefont {Katcho},\ and\ \citenamefont
  {Mingo}}]{Li-shengbte-2014}%
  \BibitemOpen
  \bibfield  {author} {\bibinfo {author} {\bibfnamefont {W.}~\bibnamefont
  {Li}}, \bibinfo {author} {\bibfnamefont {J.}~\bibnamefont {Carrete}},
  \bibinfo {author} {\bibfnamefont {N.~A.}\ \bibnamefont {Katcho}}, \ and\
  \bibinfo {author} {\bibfnamefont {N.}~\bibnamefont {Mingo}},\ }\href@noop {}
  {\bibfield  {journal} {\bibinfo  {journal} {Comput. Phys. Commun.}\ }\textbf
  {\bibinfo {volume} {185}},\ \bibinfo {pages} {1747} (\bibinfo {year}
  {2014})}\BibitemShut {NoStop}%
\bibitem [{\citenamefont {Maradudin}\ and\ \citenamefont
  {Fein}(1962)}]{Maradudin-Fein-ScatteringOfNeutrons-1962}%
  \BibitemOpen
  \bibfield  {author} {\bibinfo {author} {\bibfnamefont {A.}~\bibnamefont
  {Maradudin}}\ and\ \bibinfo {author} {\bibfnamefont {A.}~\bibnamefont
  {Fein}},\ }\href@noop {} {\bibfield  {journal} {\bibinfo  {journal} {Phys.
  Rev.}\ }\textbf {\bibinfo {volume} {128}},\ \bibinfo {pages} {2589} (\bibinfo
  {year} {1962})}\BibitemShut {NoStop}%
\bibitem [{\citenamefont {Paulatto}\ \emph {et~al.}(2015)\citenamefont
  {Paulatto}, \citenamefont {Errea}, \citenamefont {Calandra},\ and\
  \citenamefont {Mauri}}]{Paulatto-PhysRevB.91.054304_lifetimes-2015}%
  \BibitemOpen
  \bibfield  {author} {\bibinfo {author} {\bibfnamefont {L.}~\bibnamefont
  {Paulatto}}, \bibinfo {author} {\bibfnamefont {I.}~\bibnamefont {Errea}},
  \bibinfo {author} {\bibfnamefont {M.}~\bibnamefont {Calandra}}, \ and\
  \bibinfo {author} {\bibfnamefont {F.}~\bibnamefont {Mauri}},\ }\href
  {\doibase 10.1103/PhysRevB.91.054304} {\bibfield  {journal} {\bibinfo
  {journal} {Phys. Rev. B}\ }\textbf {\bibinfo {volume} {91}},\ \bibinfo
  {pages} {054304} (\bibinfo {year} {2015})}\BibitemShut {NoStop}%
\bibitem [{\citenamefont {Romero}\ \emph {et~al.}(2015)\citenamefont {Romero},
  \citenamefont {Gross}, \citenamefont {Verstraete},\ and\ \citenamefont
  {Hellman}}]{Romero-TcondPbTe-PhysRevB.91.214310-2015}%
  \BibitemOpen
  \bibfield  {author} {\bibinfo {author} {\bibfnamefont {A.~H.}\ \bibnamefont
  {Romero}}, \bibinfo {author} {\bibfnamefont {E.~K.~U.}\ \bibnamefont
  {Gross}}, \bibinfo {author} {\bibfnamefont {M.~J.}\ \bibnamefont
  {Verstraete}}, \ and\ \bibinfo {author} {\bibfnamefont {O.}~\bibnamefont
  {Hellman}},\ }\href {\doibase 10.1103/PhysRevB.91.214310} {\bibfield
  {journal} {\bibinfo  {journal} {Phys. Rev. B}\ }\textbf {\bibinfo {volume}
  {91}},\ \bibinfo {pages} {214310} (\bibinfo {year} {2015})}\BibitemShut
  {NoStop}%
\bibitem [{\citenamefont {Momma}\ and\ \citenamefont
  {Izumi}(2011)}]{Momma-Vesta-2011}%
  \BibitemOpen
  \bibfield  {author} {\bibinfo {author} {\bibfnamefont {K.}~\bibnamefont
  {Momma}}\ and\ \bibinfo {author} {\bibfnamefont {F.}~\bibnamefont {Izumi}},\
  }\href@noop {} {\bibfield  {journal} {\bibinfo  {journal} {J. Appl.
  Crystallogr.}\ }\textbf {\bibinfo {volume} {44}},\ \bibinfo {pages} {1272}
  (\bibinfo {year} {2011})}\BibitemShut {NoStop}%
\bibitem [{\citenamefont {Shevelkov}\ and\ \citenamefont
  {Kovnir}(2011)}]{Shevelkov-zintl-clath-2011}%
  \BibitemOpen
  \bibfield  {author} {\bibinfo {author} {\bibfnamefont {A.}~\bibnamefont
  {Shevelkov}}\ and\ \bibinfo {author} {\bibfnamefont {K.}~\bibnamefont
  {Kovnir}},\ }\href@noop {} {\bibfield  {journal} {\bibinfo  {journal}
  {Struct. Bond.}\ }\textbf {\bibinfo {volume} {139}},\ \bibinfo {pages} {97}
  (\bibinfo {year} {2011})}\BibitemShut {NoStop}%
\bibitem [{\citenamefont {Gatti}\ \emph {et~al.}(2003)\citenamefont {Gatti},
  \citenamefont {Bertini}, \citenamefont {Blake},\ and\ \citenamefont
  {Iversen}}]{Gatti-GuestFrameworkInt-2003}%
  \BibitemOpen
  \bibfield  {author} {\bibinfo {author} {\bibfnamefont {C.}~\bibnamefont
  {Gatti}}, \bibinfo {author} {\bibfnamefont {L.}~\bibnamefont {Bertini}},
  \bibinfo {author} {\bibfnamefont {N.~P.}\ \bibnamefont {Blake}}, \ and\
  \bibinfo {author} {\bibfnamefont {B.~B.}\ \bibnamefont {Iversen}},\
  }\href@noop {} {\bibfield  {journal} {\bibinfo  {journal} {Chem. Eur. J.}\
  }\textbf {\bibinfo {volume} {9}},\ \bibinfo {pages} {4556} (\bibinfo {year}
  {2003})}\BibitemShut {NoStop}%
\bibitem [{\citenamefont {Giannozzi}\ \emph {et~al.}(2009)\citenamefont
  {Giannozzi}, \citenamefont {Baroni}, \citenamefont {Bonini}, \citenamefont
  {Calandra}, \citenamefont {Car}, \citenamefont {Cavazzoni}, \citenamefont
  {Ceresoli}, \citenamefont {Chiarotti}, \citenamefont {Cococcioni},
  \citenamefont {Dabo}, \citenamefont {{Dal Corso}}, \citenamefont
  {de~Gironcoli}, \citenamefont {Fabris}, \citenamefont {Fratesi},
  \citenamefont {Gebauer}, \citenamefont {Gerstmann}, \citenamefont
  {Gougoussis}, \citenamefont {Kokalj}, \citenamefont {Lazzeri}, \citenamefont
  {Martin-Samos}, \citenamefont {Marzari}, \citenamefont {Mauri}, \citenamefont
  {Mazzarello}, \citenamefont {Paolini}, \citenamefont {Pasquarello},
  \citenamefont {Paulatto}, \citenamefont {Sbraccia}, \citenamefont {Scandolo},
  \citenamefont {Sclauzero}, \citenamefont {Seitsonen}, \citenamefont
  {Smogunov}, \citenamefont {Umari},\ and\ \citenamefont
  {Wentzcovitch}}]{QE-2009}%
  \BibitemOpen
  \bibfield  {author} {\bibinfo {author} {\bibfnamefont {P.}~\bibnamefont
  {Giannozzi}}, \bibinfo {author} {\bibfnamefont {S.}~\bibnamefont {Baroni}},
  \bibinfo {author} {\bibfnamefont {N.}~\bibnamefont {Bonini}}, \bibinfo
  {author} {\bibfnamefont {M.}~\bibnamefont {Calandra}}, \bibinfo {author}
  {\bibfnamefont {R.}~\bibnamefont {Car}}, \bibinfo {author} {\bibfnamefont
  {C.}~\bibnamefont {Cavazzoni}}, \bibinfo {author} {\bibfnamefont
  {D.}~\bibnamefont {Ceresoli}}, \bibinfo {author} {\bibfnamefont {G.~L.}\
  \bibnamefont {Chiarotti}}, \bibinfo {author} {\bibfnamefont {M.}~\bibnamefont
  {Cococcioni}}, \bibinfo {author} {\bibfnamefont {I.}~\bibnamefont {Dabo}},
  \bibinfo {author} {\bibfnamefont {A.}~\bibnamefont {{Dal Corso}}}, \bibinfo
  {author} {\bibfnamefont {S.}~\bibnamefont {de~Gironcoli}}, \bibinfo {author}
  {\bibfnamefont {S.}~\bibnamefont {Fabris}}, \bibinfo {author} {\bibfnamefont
  {G.}~\bibnamefont {Fratesi}}, \bibinfo {author} {\bibfnamefont
  {R.}~\bibnamefont {Gebauer}}, \bibinfo {author} {\bibfnamefont
  {U.}~\bibnamefont {Gerstmann}}, \bibinfo {author} {\bibfnamefont
  {C.}~\bibnamefont {Gougoussis}}, \bibinfo {author} {\bibfnamefont
  {A.}~\bibnamefont {Kokalj}}, \bibinfo {author} {\bibfnamefont
  {M.}~\bibnamefont {Lazzeri}}, \bibinfo {author} {\bibfnamefont
  {L.}~\bibnamefont {Martin-Samos}}, \bibinfo {author} {\bibfnamefont
  {N.}~\bibnamefont {Marzari}}, \bibinfo {author} {\bibfnamefont
  {F.}~\bibnamefont {Mauri}}, \bibinfo {author} {\bibfnamefont
  {R.}~\bibnamefont {Mazzarello}}, \bibinfo {author} {\bibfnamefont
  {S.}~\bibnamefont {Paolini}}, \bibinfo {author} {\bibfnamefont
  {A.}~\bibnamefont {Pasquarello}}, \bibinfo {author} {\bibfnamefont
  {L.}~\bibnamefont {Paulatto}}, \bibinfo {author} {\bibfnamefont
  {C.}~\bibnamefont {Sbraccia}}, \bibinfo {author} {\bibfnamefont
  {S.}~\bibnamefont {Scandolo}}, \bibinfo {author} {\bibfnamefont
  {G.}~\bibnamefont {Sclauzero}}, \bibinfo {author} {\bibfnamefont {A.~P.}\
  \bibnamefont {Seitsonen}}, \bibinfo {author} {\bibfnamefont {A.}~\bibnamefont
  {Smogunov}}, \bibinfo {author} {\bibfnamefont {P.}~\bibnamefont {Umari}}, \
  and\ \bibinfo {author} {\bibfnamefont {R.~M.}\ \bibnamefont {Wentzcovitch}},\
  }\href {http://www.quantum-espresso.org} {\bibfield  {journal} {\bibinfo
  {journal} {J. Phys.: Condens. Matter}\ }\textbf {\bibinfo {volume} {21}},\
  \bibinfo {pages} {395502} (\bibinfo {year} {2009})}\BibitemShut {NoStop}%
\bibitem [{\citenamefont {Garrity}\ \emph {et~al.}(2014)\citenamefont
  {Garrity}, \citenamefont {Bennett}, \citenamefont {Rabe},\ and\ \citenamefont
  {Vanderbilt}}]{Garrity-pseudopotentials-2014}%
  \BibitemOpen
  \bibfield  {author} {\bibinfo {author} {\bibfnamefont {K.~F.}\ \bibnamefont
  {Garrity}}, \bibinfo {author} {\bibfnamefont {J.~W.}\ \bibnamefont
  {Bennett}}, \bibinfo {author} {\bibfnamefont {K.~M.}\ \bibnamefont {Rabe}}, \
  and\ \bibinfo {author} {\bibfnamefont {D.}~\bibnamefont {Vanderbilt}},\
  }\href@noop {} {\bibfield  {journal} {\bibinfo  {journal} {Comput. Mater.
  Sci.}\ }\textbf {\bibinfo {volume} {81}},\ \bibinfo {pages} {446} (\bibinfo
  {year} {2014})}\BibitemShut {NoStop}%
\bibitem [{\citenamefont {Perdew}\ \emph {et~al.}(1996)\citenamefont {Perdew},
  \citenamefont {Burke},\ and\ \citenamefont
  {Ernzerhof}}]{Perdew-generalized-1996}%
  \BibitemOpen
  \bibfield  {author} {\bibinfo {author} {\bibfnamefont {J.~P.}\ \bibnamefont
  {Perdew}}, \bibinfo {author} {\bibfnamefont {K.}~\bibnamefont {Burke}}, \
  and\ \bibinfo {author} {\bibfnamefont {M.}~\bibnamefont {Ernzerhof}},\
  }\href@noop {} {\bibfield  {journal} {\bibinfo  {journal} {Phys. Rev. Lett.}\
  }\textbf {\bibinfo {volume} {77}},\ \bibinfo {pages} {3865} (\bibinfo {year}
  {1996})}\BibitemShut {NoStop}%
\bibitem [{\citenamefont {Li}\ \emph {et~al.}(2012{\natexlab{a}})\citenamefont
  {Li}, \citenamefont {Mingo}, \citenamefont {Lindsay}, \citenamefont {Broido},
  \citenamefont {Stewart},\ and\ \citenamefont
  {Katcho}}]{Li-Gaussian_PhysRevB.85.195436-2012}%
  \BibitemOpen
  \bibfield  {author} {\bibinfo {author} {\bibfnamefont {W.}~\bibnamefont
  {Li}}, \bibinfo {author} {\bibfnamefont {N.}~\bibnamefont {Mingo}}, \bibinfo
  {author} {\bibfnamefont {L.}~\bibnamefont {Lindsay}}, \bibinfo {author}
  {\bibfnamefont {D.~A.}\ \bibnamefont {Broido}}, \bibinfo {author}
  {\bibfnamefont {D.~A.}\ \bibnamefont {Stewart}}, \ and\ \bibinfo {author}
  {\bibfnamefont {N.~A.}\ \bibnamefont {Katcho}},\ }\href@noop {} {\bibfield
  {journal} {\bibinfo  {journal} {Phys. Rev. B}\ }\textbf {\bibinfo {volume}
  {85}},\ \bibinfo {pages} {195436} (\bibinfo {year}
  {2012}{\natexlab{a}})}\BibitemShut {NoStop}%
\bibitem [{\citenamefont {Li}\ \emph {et~al.}(2012{\natexlab{b}})\citenamefont
  {Li}, \citenamefont {Lindsay}, \citenamefont {Broido}, \citenamefont
  {Stewart},\ and\ \citenamefont
  {Mingo}}]{Li-ThirdOrderPy_PhysRevB.86.174307-2012}%
  \BibitemOpen
  \bibfield  {author} {\bibinfo {author} {\bibfnamefont {W.}~\bibnamefont
  {Li}}, \bibinfo {author} {\bibfnamefont {L.}~\bibnamefont {Lindsay}},
  \bibinfo {author} {\bibfnamefont {D.~A.}\ \bibnamefont {Broido}}, \bibinfo
  {author} {\bibfnamefont {D.~A.}\ \bibnamefont {Stewart}}, \ and\ \bibinfo
  {author} {\bibfnamefont {N.}~\bibnamefont {Mingo}},\ }\href@noop {}
  {\bibfield  {journal} {\bibinfo  {journal} {Phys. Rev. B}\ }\textbf {\bibinfo
  {volume} {86}},\ \bibinfo {pages} {174307} (\bibinfo {year}
  {2012}{\natexlab{b}})}\BibitemShut {NoStop}%
\bibitem [{\citenamefont {Bell}\ \emph {et~al.}(1970)\citenamefont {Bell},
  \citenamefont {Dean},\ and\ \citenamefont
  {Hibbins-Butler}}]{Bell-Localization-0022-3719-3-10-013-1970}%
  \BibitemOpen
  \bibfield  {author} {\bibinfo {author} {\bibfnamefont {R.~J.}\ \bibnamefont
  {Bell}}, \bibinfo {author} {\bibfnamefont {P.}~\bibnamefont {Dean}}, \ and\
  \bibinfo {author} {\bibfnamefont {D.~C.}\ \bibnamefont {Hibbins-Butler}},\
  }\href {http://stacks.iop.org/0022-3719/3/i=10/a=013} {\bibfield  {journal}
  {\bibinfo  {journal} {J. Phys. C}\ }\textbf {\bibinfo {volume} {3}},\
  \bibinfo {pages} {2111} (\bibinfo {year} {1970})}\BibitemShut {NoStop}%
\bibitem [{\citenamefont {Hafner}\ and\ \citenamefont
  {Krajci}(1993)}]{Hafner-PropagatingExQuasiCryst-1993}%
  \BibitemOpen
  \bibfield  {author} {\bibinfo {author} {\bibfnamefont {J.}~\bibnamefont
  {Hafner}}\ and\ \bibinfo {author} {\bibfnamefont {M.}~\bibnamefont
  {Krajci}},\ }\href@noop {} {\bibfield  {journal} {\bibinfo  {journal} {J.
  Phys.: Condens. Matter}\ }\textbf {\bibinfo {volume} {5}},\ \bibinfo {pages}
  {2489} (\bibinfo {year} {1993})}\BibitemShut {NoStop}%
\bibitem [{\citenamefont {Barron}\ \emph {et~al.}(1974)\citenamefont {Barron},
  \citenamefont {Klein}, \citenamefont {Horton},\ and\ \citenamefont
  {Maradudin}}]{Barron-DynamicalPropertiesOfSolids-1974}%
  \BibitemOpen
  \bibfield  {author} {\bibinfo {author} {\bibfnamefont {T.}~\bibnamefont
  {Barron}}, \bibinfo {author} {\bibfnamefont {M.}~\bibnamefont {Klein}},
  \bibinfo {author} {\bibfnamefont {G.}~\bibnamefont {Horton}}, \ and\ \bibinfo
  {author} {\bibfnamefont {A.}~\bibnamefont {Maradudin}},\ }\href@noop {}
  {\emph {\bibinfo {title} {Perturbation Theory of Anharmonic Crystals}}},\
  Vol.~\bibinfo {volume} {1}\ (\bibinfo  {publisher} {Amsterdam:
  North-Holland},\ \bibinfo {year} {1974})\ pp.\ \bibinfo {pages}
  {391--450}\BibitemShut {NoStop}%
\bibitem [{\citenamefont {Kennedy}\ and\ \citenamefont
  {White}(2005)}]{Kennedy-NTE-thermal-cond-ZrW2O8-2005}%
  \BibitemOpen
  \bibfield  {author} {\bibinfo {author} {\bibfnamefont {C.~A.}\ \bibnamefont
  {Kennedy}}\ and\ \bibinfo {author} {\bibfnamefont {M.~A.}\ \bibnamefont
  {White}},\ }\href {\doibase http://dx.doi.org/10.1016/j.ssc.2005.01.031}
  {\bibfield  {journal} {\bibinfo  {journal} {Solid State Commun.}\ }\textbf
  {\bibinfo {volume} {134}},\ \bibinfo {pages} {271 } (\bibinfo {year}
  {2005})}\BibitemShut {NoStop}%
\bibitem [{\citenamefont {Kennedy}\ \emph {et~al.}(2007)\citenamefont
  {Kennedy}, \citenamefont {White}, \citenamefont {Wilkinson},\ and\
  \citenamefont {Varga}}]{Kennedy-NTE-thermal-cond-HfMo2O8-2007}%
  \BibitemOpen
  \bibfield  {author} {\bibinfo {author} {\bibfnamefont {C.~A.}\ \bibnamefont
  {Kennedy}}, \bibinfo {author} {\bibfnamefont {M.~A.}\ \bibnamefont {White}},
  \bibinfo {author} {\bibfnamefont {A.~P.}\ \bibnamefont {Wilkinson}}, \ and\
  \bibinfo {author} {\bibfnamefont {T.}~\bibnamefont {Varga}},\ }\href
  {\doibase 10.1063/1.2721860} {\bibfield  {journal} {\bibinfo  {journal}
  {Appl. Phys. Lett.}\ }\textbf {\bibinfo {volume} {90}},\ \bibinfo {eid}
  {151906} (\bibinfo {year} {2007})}\BibitemShut {NoStop}%
\bibitem [{\citenamefont
  {Wallace}(1972)}]{Wallace-ThermodynamicsOfCrystals-1972}%
  \BibitemOpen
  \bibfield  {author} {\bibinfo {author} {\bibfnamefont {D.}~\bibnamefont
  {Wallace}},\ }\href@noop {} {\emph {\bibinfo {title} {Thermodynamics of
  Crystals}}}\ (\bibinfo  {publisher} {John Wiley \& Sons, New York},\ \bibinfo
  {year} {1972})\BibitemShut {NoStop}%
\bibitem [{\citenamefont {H{\"a}rk{\"o}nen}\ and\ \citenamefont
  {Karttunen}(2014)}]{Harkonen-NTE-2014}%
  \BibitemOpen
  \bibfield  {author} {\bibinfo {author} {\bibfnamefont {V.~J.}\ \bibnamefont
  {H{\"a}rk{\"o}nen}}\ and\ \bibinfo {author} {\bibfnamefont {A.~J.}\
  \bibnamefont {Karttunen}},\ }\href@noop {} {\bibfield  {journal} {\bibinfo
  {journal} {Phys. Rev. B}\ }\textbf {\bibinfo {volume} {89}},\ \bibinfo
  {pages} {024305} (\bibinfo {year} {2014})}\BibitemShut {NoStop}%
\bibitem [{\citenamefont {Li}\ and\ \citenamefont
  {Mingo}(2015)}]{Li-UltraLowLatCond-PhysRevB.91.144304-2015}%
  \BibitemOpen
  \bibfield  {author} {\bibinfo {author} {\bibfnamefont {W.}~\bibnamefont
  {Li}}\ and\ \bibinfo {author} {\bibfnamefont {N.}~\bibnamefont {Mingo}},\
  }\href {\doibase 10.1103/PhysRevB.91.144304} {\bibfield  {journal} {\bibinfo
  {journal} {Phys. Rev. B}\ }\textbf {\bibinfo {volume} {91}},\ \bibinfo
  {pages} {144304} (\bibinfo {year} {2015})}\BibitemShut {NoStop}%
\bibitem [{\citenamefont {Chakoumakos}\ \emph {et~al.}(2000)\citenamefont
  {Chakoumakos}, \citenamefont {Sales}, \citenamefont {Mandrus},\ and\
  \citenamefont {Nolas}}]{Chakoumakos-StructuralDisordLatCond-2000}%
  \BibitemOpen
  \bibfield  {author} {\bibinfo {author} {\bibfnamefont {B.}~\bibnamefont
  {Chakoumakos}}, \bibinfo {author} {\bibfnamefont {B.}~\bibnamefont {Sales}},
  \bibinfo {author} {\bibfnamefont {D.}~\bibnamefont {Mandrus}}, \ and\
  \bibinfo {author} {\bibfnamefont {G.}~\bibnamefont {Nolas}},\ }\href@noop {}
  {\bibfield  {journal} {\bibinfo  {journal} {J. Alloys Compd.}\ }\textbf
  {\bibinfo {volume} {296}},\ \bibinfo {pages} {80} (\bibinfo {year}
  {2000})}\BibitemShut {NoStop}%
\bibitem [{\citenamefont {Chakoumakos}\ \emph {et~al.}(2001)\citenamefont
  {Chakoumakos}, \citenamefont {Sales},\ and\ \citenamefont
  {Mandrus}}]{Chakoumakos-StructuralDisordClath-2001}%
  \BibitemOpen
  \bibfield  {author} {\bibinfo {author} {\bibfnamefont {B.}~\bibnamefont
  {Chakoumakos}}, \bibinfo {author} {\bibfnamefont {B.}~\bibnamefont {Sales}},
  \ and\ \bibinfo {author} {\bibfnamefont {D.}~\bibnamefont {Mandrus}},\
  }\href@noop {} {\bibfield  {journal} {\bibinfo  {journal} {J. Alloys Compd.}\
  }\textbf {\bibinfo {volume} {322}},\ \bibinfo {pages} {127} (\bibinfo {year}
  {2001})}\BibitemShut {NoStop}%
\end{thebibliography}%
\end{document}